\documentclass[aps,prx,twocolumn,amsmath,10pt,floatfix]{revtex4-2}

\usepackage{amsthm}
\usepackage{needspace}
\usepackage{appendix}
\newtheorem{theorem}{Theorem}[section]
\newtheorem{corollary}[theorem]{Corollary}

\usepackage{graphicx}
\graphicspath{{./}}
\usepackage[normalem]{ulem}
\usepackage{xcolor}
\usepackage[utf8]{inputenc}
\usepackage{comment}
\usepackage[ruled,linesnumbered]{algorithm2e}
\DeclareMathSizes{8}{7}{7}{7}

\usepackage{textgreek}
\usepackage[colorlinks, linkcolor=blue, citecolor=blue]{hyperref}
\usepackage{gensymb}
\usepackage[compat=1.1.0]{tikz-feynman}
\usepackage{tikz}
\tikzset{
pattern size/.store in=\mcSize, 
pattern size = 5pt,
pattern thickness/.store in=\mcThickness, 
pattern thickness = 0.3pt,
pattern radius/.store in=\mcRadius, 
pattern radius = 1pt}

\usepackage{bbold}
\usepackage{mathtools}

\newcommand{\be}{\begin{equation}}
\newcommand{\ee}{\end{equation}}
\newcommand{\beq}{\begin{eqnarray}}
\newcommand{\eeq}{\end{eqnarray}}
\newcommand{\ba}{\[\begin{aligned}}
\newcommand{\ea}{\end{aligned}\]}
\newcommand{\bal}{\begin{aligned}}
\newcommand{\eal}{\end{aligned}}

\newcommand{\Tr}{{\rm Tr\,}}

\renewcommand{\epsilon}{\varepsilon}
\renewcommand{\dag}{\dagger}

\def \ket#1{{|#1\rangle}}
\def \bra#1{{\langle#1|}}

\def \tn{\textnormal}

\def \ba{\begin{align*}}
\def \ea{\end{align*}}

\newcounter{indice}

\begin{document}
\title{Quantum Cooling by Classical Shadows}
\author{Xuepeng Wang}\email{xw577@cornell.edu}
\author{Debanjan Chowdhury}\email{debanjanchowdhury@cornell.edu}
\affiliation{Department of Physics, Cornell University, Ithaca, New York 14853, USA.}
\begin{abstract}
Cooling to ground states of many-body Hamiltonians by imaginary-time evolution is central to quantum simulation, but its non-unitary aspect prevents direct implementation on a quantum computer. Existing routes restore it by importing structure from outside the intrinsic quantum dynamics, e.g. by parameter estimation on many copies or recursive compilation. Here, we show that a single randomized snapshot of the quantum many-body system is sufficient to steer the state along the cooling direction on average. Our algorithm, {\it shadow chain}, uses randomized measurements to collapse the state to its classical record, which fixes the next unitary update. Snapshots across steps form a trajectory that is a quantum-executable sample of an auxiliary-field path integral; fluctuations among trajectories are exponentially suppressed in system size. With the randomness supplied intrinsically by the Born measurement rule, the trajectories sample themselves rather than going through external accept-reject steps. The resulting algorithm is ansatz-free and exact upon extrapolation to the continuous limit, with the exponential cost confined only to system size rather than temperature.

\end{abstract}

\maketitle

\section{Introduction}

The Feynman path-integral is a central organizing principle in quantum mechanics~\cite{Feynman1948}. Its Euclidean analytic  continuation~\cite{Kac1949} provides a foundational framework for evaluating the equilibrium thermal and ground-state properties of quantum many-body systems, and is a defining problem of condensed matter physics and quantum chemistry. 
Beneath the diversity of methods developed to tackle this problem lies the single operation of imaginary time propagation, $e^{-\beta h}$, which cools the system for a many-body Hamiltonian $h$ and inverse temperature $\beta$. The modern implementations of such imaginary time evolution rests on two pillars. Classically, determinant and world-line quantum Monte Carlo (QMC) stochastically sample the Euclidean path integral~\cite{BSS1981,AssaadEvertz2008,SandvikKurkijarvi1991} and are obstructed by the infamous sign problem~\cite{Loh1990,TroyerWiese2005}.
Quantum algorithms instead largely pursue a complementary route, seeking to coherently evolve in imaginary time, and are limited by overheads in circuit-depth and success probability~\cite{GilyenSuLowWiebe2019QSVT,Gluza2026DBQITE}.

A key observation is that both limitations share the same conceptual origin; they import structure from outside the quantum dynamics, signaling a failure of \emph{intrinsicness}. The sign problem arises from interpreting a non-positive-definite path integral measure as a probability distribution; imaginary-time evolution (ITE) is non-unitary, and a unitary device cannot execute it directly.
In each case, the missing structure as a remedy must be imported from outside, through reweighting or constraint approximations in QMC~\cite{ZhangKrakauer2003}, and through ancillas, post-selection, or classical optimization in quantum algorithms~\cite{PoulinWocjan2009Gibbs,Motta2020QITE,McArdle2019VarQITE}.
The associated cost in the worst case can lead to an exponential in both $\beta$ and $L$ (system size): statistical variance in the former, circuit depth or success probability in the latter. The two limitations are not identical since the sign problem is representation-dependent and can be removed by a fortunate choice of basis, whereas nonunitarity is a structural property of $e^{-\beta H}$ that no change of representation repairs.

\begin{figure*}[htb!]
\centering
\includegraphics[width=18cm]{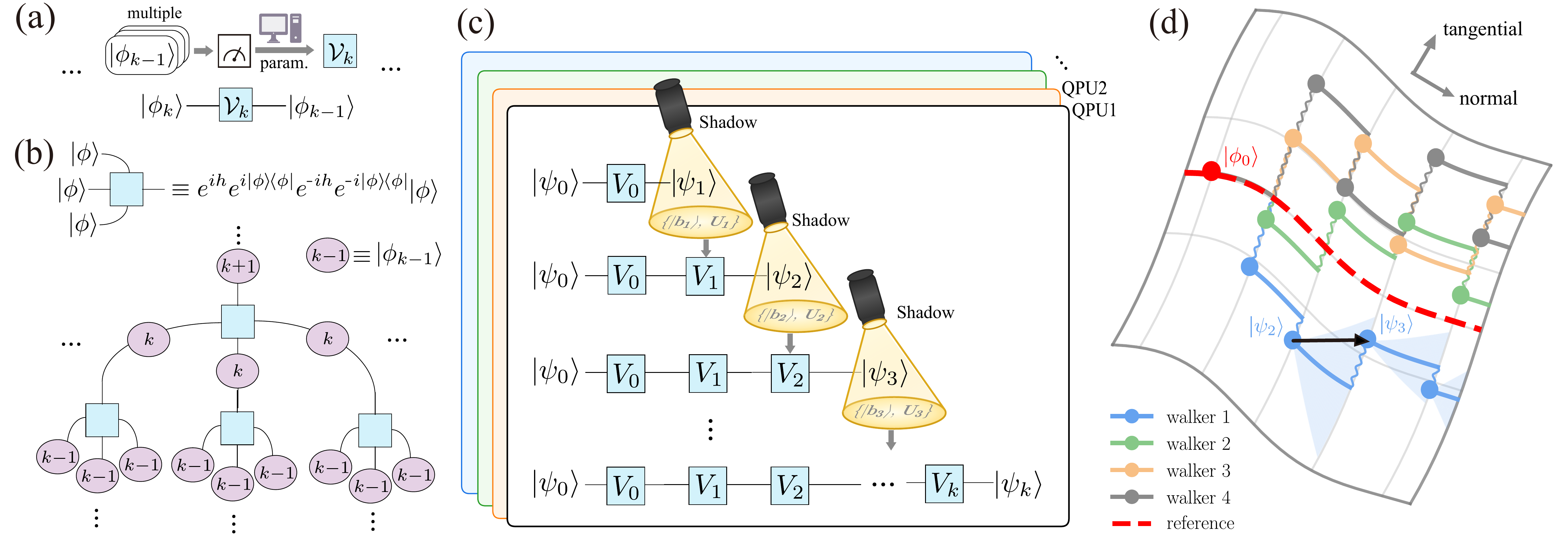}
\caption{\label{Fig:schematics}
{\bf Imaginary-time evolution and shadow-chain algorithm.}
{\bf (a)} ITE with multi-copy parameter estimation to determine the next update. 
{\bf (b)} Recursive double-bracket tree: blue boxes denote the double-bracket evolution; each consumes two corresponding states $\ket{\phi}$ and act on a freshly prepared one. Purple ellipses denote states at successive time steps.
{\bf (c)} Schematic of shadow chain algorithm. The known initial state $|\psi_0\rangle$ makes $V_0$ deterministic.
A single shadow measurement of $|\psi_1\rangle$ produces the classical record $\omega_1=\{b_1,U_1\}$ (ellipse), fixing the next shadow evolution $V_1$.
Because the measurement collapses $\ket{\psi_1}$, the walker is restarted from $\ket{\psi_0}$ and replayed through $V_0$ and $V_1$ to reach $\ket{\psi_2}$, whose snapshot $\omega_2=\{b_2,U_2\}$ fixes $V_2$; the process repeats until the target inverse temperature is reached.
Distinct trajectories are statistically independent and can be sampled in parallel
on separate quantum processing units (QPUs).
{\bf (d)} Stochastic trajectories and imaginary-time path-integral sampling. 
Gray lines along the normal direction show a family of cooling flows starting from different initial states; the red dashed line marks the reference trajectory $|\phi_0\rangle$.
Colored circles denote individual state trajectories sampled on a discrete imaginary-time grid. As an example, the update $V_2$ mapping $\ket{\psi_{2}}$ to $\ket{\psi_{3}}$ separates into conditional drift (colored solid lines) and measurement-induced fluctuation (colored shading) through the Doob decomposition (see Appendix).
}
\end{figure*}

Extensive efforts have been made to overcome these limitations~\cite{ChowdhurySomma2017Gibbs,HuoLi2023MonteCarloITE,Arulandu2026TITE,Magann2022FALQON,Magann2023RandomizedAdaptive,WiersemaKilloran2023RiemannianFlow,Temme2011Metropolis,YungAspuruGuzik2012Metropolis,ChenKastoryanoBrandaoGilyen2023ThermalPrep,ChenKastoryanoGilyen2024GibbsSampler,DingLiLin2025KMSGibbsSampler,LinTong2020NearOptimal,DongLinTong2022Eigenvalue,Bakshi2024HighTemperatureGibbs,Layden2023QuantumEnhancedMCMC,Wan2023MatchgateShadows,Zhang2025QuantumComputingQMC,Mao2023MeasurementITE,Schwartzman2026MultiCopyITE,Kolotouros2025RMITE}.
Quantum imaginary-time evolution (QITE) and its variational variants~\cite{Motta2020QITE,McArdle2019VarQITE} replace nonunitary imaginary-time steps by optimization-based state-dependent unitary dynamics (Fig.~\ref{Fig:schematics}a), while double-bracket constructions use recursively defined gates~\cite{Gluza2026DBQITE,Gluza2024Diagonalization} or externally specified stochastic protocols~\cite{McMahonPervezArenz2026,McMahonPervezArenz_arxiv} (Fig.~\ref{Fig:schematics}b).
Other proposals use quantum hardware to assist Monte Carlo by supplying trial states~\cite{Huggins2022NatureQCQMC}, performing coherent sampling~\cite{dq2mc}, or sampling Euclidean path-integral with externally specified weights~\cite{Clemente2020ThermalAverages,Yamamoto2022QuantumSampling,Son2025QDP}.
All of these remedies, however, either import structure from outside or require variational optimization~\cite{Yuan2019Theory}, in-situ parameter estimation~\cite{Suzuki2025DBQSP}, recursive compilation, or post-selection~\cite{KolotourosGarciaPatron2026PDBQITE}, and hence are not intrinsic to the underlying quantum hardware.
 
These developments sharpen three outstanding questions. 
First, can a stochastic quantum algorithm realize the Euclidean path integral {\it intrinsically} on the space of quantum states, with randomness from the qubits themselves, while avoiding in-situ parameter estimation using multiple copies of intermediate states?
Second, once realized stochastically, how much information does a \emph{single} path configuration contain, i.e. is trajectory averaging a necessity, or does a typical trajectory already carry the trajectory-averaged result? 
Third, if an exponential cost is unavoidable, must it compound with the inverse temperature, or can it be confined entirely to the system size, thereby improving the asymptotic performance at low temperatures?
Using classical shadow tomography~\cite{HuangKuengPreskill2020,Elben2023Toolbox,Aaronson2020ShadowTomography} and a thermofield double (TFD) purification,
we answer these questions affirmatively by introducing the \emph{shadow chain} (Fig.~\ref{Fig:schematics}c), a shadow-guided Markov chain that stochastically realizes the Euclidean path-integral.

We answer the first question by demonstrating that a {\it single} shadow measurement is sufficient to steer the evolution forward (Fig.~\ref{Fig:schematics}d), with randomness supplied by the Born rule; no ancilla, in-situ estimation, variational optimization, or rejection steps are required. 
Each instance is an intrinsic Born-rule-driven random walker on the manifold of pure states.  
Interestingly, each snapshot acts as an auxiliary field, and the conditional drift follows imaginary-time cooling up to finite-step corrections.
The full shadow history constitutes an auxiliary-field configuration sampled from a Born-weighted path measure. Trajectory averaging therefore is an auxiliary-field path integral, with each random walker directly executable on quantum hardware.

We address the second question by developing a stochastic theory of the shadow chain dynamics using the Doob decomposition~\cite{Williams1991}, which separates each random update into a deterministic drift and fluctuations about it (Fig.~\ref{Fig:schematics}d).   
Because averaging random updates does not commute with nonlinear evolution on the curved quantum state manifold and each gate sequence only approximates the ideal imaginary-time flow up to small corrections,
observable estimation can have systematic error. 
We show that the total systematic error grows at most polynomially in the inverse temperature $\beta$ and the system size $L$, giving {\it unbiased exact results} upon extrapolation.
Interestingly, the chain concentrates \emph{exponentially}, i.e. the trajectory-to-trajectory fluctuations are exponentially suppressed in system size, so a typical walker returns the trajectory-averaged answer at sufficiently large system size.
We refer to this as \emph{trajectory typicality}, which is an extension to the canonical typicality~\cite{Popescu2006,Goldstein2006,SugiuraShimizu2012TPQ,SugiuraShimizu2013TPQ,Coopmans2023PureThermalShadows,White2009METTS,StoudenmireWhite2010METTS} and focuses on the concentration of the stochastically prepared states at each intermediate step as a sequence of canonical typicality.

Finally, we answer the third question by noting that the shadow measurements are classically recorded and cut off the recursive tree structure.
Repeated implementation of the shadow evolution forms a chain structure with the exponential overhead confined \emph{only} to the system size rather than compounding with inverse temperature. However, the price associated with this operation is the depolarizing factor of the classical-shadow channel,which suppresses the Trotter step-size by the Hilbert-space dimension. 

The shadow chain is a framework that realizes the Euclidean path integral intrinsically on qubits with several advantages over conventional approaches. Auxiliary-field QMC introduces auxiliary fields by hand and samples configurations through explicit updating rules such as Metropolis-Hastings. The shadow chain requires neither. The ``auxiliary" shadows come from measurements rather than being added externally, and the sampling is done by the qubits themselves.
Moreover, individual trajectories are typical, so for large enough systems a {\it single} trajectory already reproduces the ensemble average. 
Importantly, a single run also gives access to a whole range of temperatures. Replaying its classically recorded history prepares the states at {\it every} intermediate temperature.
Snapshots from independent trajectories can also be combined using median-of-means estimation, with sample complexity logarithmic in the number of observables and an exponential prefactor in system size from the global-shadow variance.

\section{Shadow chain}
Let us briefly review the ITE cooling flow in the double-bracket formalism~\cite{Brockett1991DoubleBracket,Wegner1994FlowEquations}, with $\ket{\phi_\tau}$ denoting the deterministic (reference) state.
For a pure state, Trotterized imaginary-time evolution can be generated by the state-dependent commutator $[\rho_{\phi_\tau},h]$ as $\ket{\phi_{\tau+\Delta\tau}} = e^{\Delta\tau[\rho_{\phi_\tau},h]}\ket{\phi_\tau}$ (see Appendix), where $\rho_{\phi_\tau}\equiv\ket{\phi_\tau}\bra{\phi_\tau}$ is the instantaneous density matrix, $h$ is the Hamiltonian, and $[\rho_{\phi_\tau},h]$ corresponds to a Riemannian gradient descent on the pure state manifold~\cite{Gluza2026DBQITE}.
Here the pure-state manifold is the projective Hilbert space $\mathbb{C}P^{d-1}$, equipped with the Fubini-Study metric, with respect to which the double-bracket flow follows the direction of steepest descent of the energy expectation value~\cite{BrodyHughston2001GeometricQM,Zander2025RiemannianDBQITE}.
Since the generator depends on the instantaneous state $|\phi_\tau\rangle$, each update requires repeated preparation of $|\phi_\tau\rangle$ through all previous steps, unfolding as a recursion tree whose depth grows exponentially with the step count (Fig.~\ref{Fig:schematics}b).

The shadow chain removes this recursion by replacing the deterministic reference state $\rho_{\phi_\tau}$ in the generator with a {\it single} classical shadow of it. 
Operationally, one measurement record fixes each update through single-shot, measurement-conditioned feedforward, without estimating state-dependent parameters from an ensemble. Let $\ket{\psi_k}$ denote the state prepared stochastically by the shadow chain.
At step $k$, we apply a random unitary $U_k^{\dag}$ drawn from the Haar ensemble 
to the current state $\ket{\psi_k}$ and measure in the computational basis, obtaining the outcome $\ket{b_k}$ with Born probability,  $p(b_k|U_k)=|\bra{b_k}U_k^\dagger\ket{\psi_k}|^2$. The resulting snapshot as a rank-one projector $P_k$ and the corresponding shadow evolution operator $V_k$ are defined as
\begin{subequations}
\label{eq:shadow_evolve}
\beq
\label{eq:shadow_def_p}
P_k &\equiv& U_k\ket{b_k}\bra{b_k}U_k^\dagger ,
\\
\label{eq:shadow_evolve_b}
V_k &=&
e^{i\alpha h}e^{i\theta P_k}e^{-i\alpha h}e^{-i\theta P_k}; ~~ \Delta\tau = \alpha\theta~~,
\eeq
\end{subequations}
where $V_{k}$ is an approximation to $e^{\alpha\theta[P_k,h]}$ up to $O(\alpha^2||h||_{\tn{op}}^2\theta,\alpha||h||_{\tn{op}}\theta^2)$ corrections. Here, $||h||_{\tn{op}}$ denotes the operator norm of $h$; $\alpha||h||_{\tn{op}}$ and $\theta$ are assumed to be the small dimensionless parameters. 
Shadow evolution then updates the state as $\ket{\psi_{k+1}}=V_k\ket{\psi_k}$.
Thus no separate estimation stage intervenes between measurement and evolution; the shadow record $\omega_k \equiv \{ U_k, \ket{b_k} \}$ directly fixes the shadow evolution operator $V_k$.

\begin{figure*}[htb!]
\centering
\includegraphics[width=\textwidth]{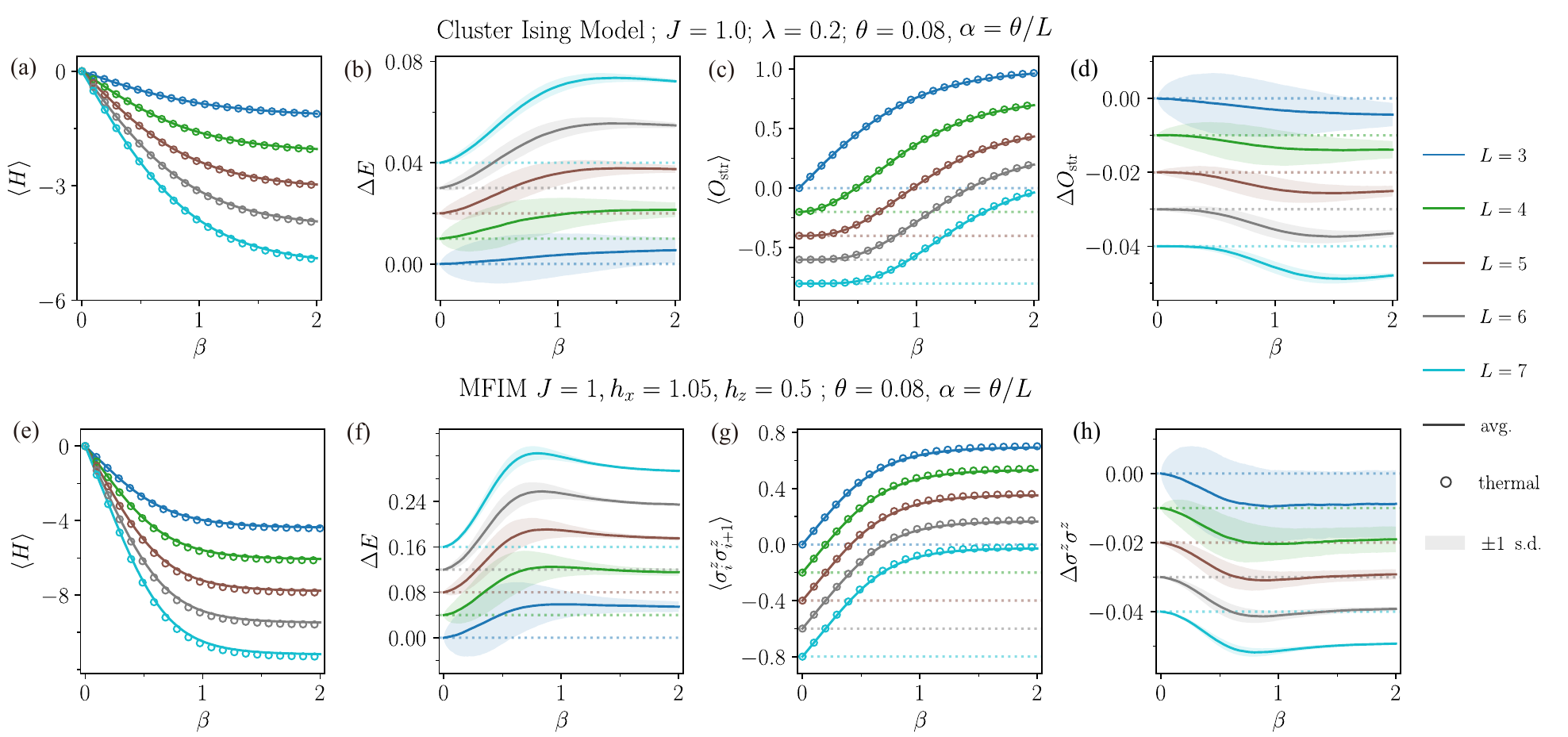}
\caption{\label{Fig:shadow_result}\textbf{Thermal observable estimation using shadow chain.} Results for
\textbf{(a)--(d)}, Cluster Ising chain, and \textbf{(e)--(h)}, mixed-field Ising chain, using global Haar randomness, with system sizes ranging from $L=3$ to $L=7$ and $\alpha JL=\theta=0.08$.
Solid lines are trajectory averages (see Supplementary Material for a discussion of required number of trajectory), open circles are exact thermal (Gibbs) expectation values, and shaded regions denote the trajectory-to-trajectory standard deviation.
Panels \textbf{(a)} and \textbf{(e)} show the energy as a function of $\beta$; \textbf{(b)} and \textbf{(f)} show $\Delta E$, the difference from the exact thermal reference. In \textbf{(c)}, we show the string operator; in \textbf{(g)}, we show the longitudinal nearest-neighbor correlations averaged over site-$i$; in \textbf{(d)} and \textbf{(h)}, their difference from the exact thermal reference.
Constant vertical offsets are added to the curves in \textbf{(b)}--\textbf{(d)}, and \textbf{(f)}--\textbf{(h)} for visual clarity (dotted baseline mark the zero for each offset). A systematic suppression of the trajectory fluctuation is observed with increasing system size. 
}
\end{figure*}

The measurement collapses $\ket{\psi_k}$, so a fresh copy of $\ket{\psi_k}$ is needed to reach $\ket{\psi_{k+1}}$. 
This is where the shadow record earns its keep;
because the shadow information $\omega_k \equiv \{ U_k, \ket{b_k} \}$ of all previous steps is recorded classically, re-preparing $\ket{\psi_k}$ only requires replaying the determined sequence $\omega_0,\dots,\omega_{k-1}$ and thus does not lead to any recursive definition or tree structure. Repeated implementation of the shadow evolution forms a chain structure (Fig.~\ref{Fig:schematics}c), since each shadow evolution $V_k$ consumes only a single snapshot.
The randomness in Eq.~\eqref{eq:shadow_evolve} is not introduced by hand, as it is the Born-rule collapse of the measurement that produces the snapshot.
Averaging $P_k$ over the Born-weighted measurement channel at fixed $\ket{\psi_k}$ gives~\cite{HuangKuengPreskill2020}  $\mathbb{E}[P_k|\rho_{\psi_k}]=(\rho_{\psi_k}+\mathbb{1})/(d+1)$, where $d$ is
the Hilbert space dimension of the register carrying the shadow (see Appendix). Thus, the single-step shadow evolution in Eq.~\eqref{eq:shadow_evolve_b} conditionally recovers the {\it exact} ITE flow after averaging, with a renormalized effective Trotter step-size $\Delta\tau_{\tn{eff}}$, defined as
$
\Delta\tau_{\tn{eff}}
=\alpha\theta/(d+1).
$
Fig.~\ref{Fig:schematics}d schematically illustrates the stochastic trajectories and their connection to random walkers.

\section{Thermal observable estimation}

Thermal observables can be estimated by applying the shadow chain to a purification of the thermal density matrix. A natural choice for such purification is via the TFD state, obtained by introducing an ancillary copy of the system qubits,  
\be
\label{eq:def_tfd}
\ket{\phi_\beta}=\frac{1}{\sqrt{\mathcal{Z}(\beta)}}\sum_{i=1}^{N}e^{-\beta H/2}\ket{i}_{\rm sys}\otimes\ket{i}_{\rm anc},
\ee
where $\beta$ is the inverse temperature and $\mathcal{Z}(\beta)=\Tr e^{-\beta H}$ is the partition function. The TFD state in Eq.~\eqref{eq:def_tfd} corresponds to applying continuous ITE cooling flow introduced earlier to the infinite-temperature state, $\ket{\phi_{\beta=0}}$, up to Trotter errors.

Tracing out the ancilla in Eq.~\eqref{eq:def_tfd} gives the Gibbs state, and hence the thermal expectation of an observable $O$ is
$
\langle O\rangle_\beta=\bra{\phi_\beta}O_{\rm sys}\otimes\mathbb{1}_{\rm anc}\ket{\phi_\beta}.
$
 To benchmark the shadow chain protocol, we initialize the chain at $\ket{\psi_0}=\ket{\phi_{\beta=0}}$. The shadow evolution defined in Eq.~\eqref{eq:shadow_evolve} is then applied repeatedly to evolve the state until reaching the target inverse temperature $\beta=2k\eta$, where $\eta=\alpha\theta/(N^2+1)$ for the doubled register and with $h\equiv H \otimes \mathbb{1}_{\rm anc}$. Using the classically-stored shadow history of each random walker $a$, we rerun the full protocol to reconstruct the corresponding pure state $\ket{\psi_\beta^{(a)}}$ and estimate the thermal observables from the trajectory average
\be
\overline{O}_\beta=\frac{1}{N_{\rm traj}}\sum_{a=1}^{N_{\rm traj}}\bra{\psi_\beta^{(a)}}O_{\rm sys}\otimes\mathbb{1}_{\rm anc}\ket{\psi_\beta^{(a)}},
\ee
where $N_{\rm traj}$ is the number of independent trajectories.

We benchmark the shadow chain on two Hamiltonians: the cluster Ising model (CIM)~\cite{Smacchia2011ClusterIsing} and the mixed-field Ising model (MFIM)~\cite{chaotic_ising,PhysRevLett.106.050405} model with open boundary condition, defined as
\begin{subequations}
\beq
H_{\mathrm{CIM}}
&=&
-J\sum_{i=2}^{L-1}\sigma_{i-1}^x\sigma_i^z\sigma_{i+1}^x
+\lambda\sum_{i=1}^{L-1}\sigma_i^y\sigma_{i+1}^y,
\\
H_{\mathrm{MFIM}}
&=&
-J\sum_{i=1}^{L-1}\sigma_i^z\sigma_{i+1}^z- h_x \sum_{i=1}^{L}\sigma_i^x - h_z \sum_{i=1}^{L}\sigma_i^z,~~~~
\eeq
\end{subequations}
where $\sigma_i^x,\sigma_i^y,\sigma_i^z$ are Pauli operators on site $i$, and $J=1$ is fixed throughout the benchmark. Further results on the MFIM with other parameters, as well as the critical transverse-field Ising model, and the antiferromagnetic Heisenberg chain can be found in the supplementary materials \cite{si}. These benchmarks demonstrate the universal applicability of the shadow chain for many-body systems, regardless of any special integrability considerations or excitations that are restricted to be local. For the CIM model, the expectation value of energy tracks the exact thermal reference [Fig.~\ref{Fig:shadow_result}(a)], with a controlled deviation [Fig.~\ref{Fig:shadow_result}(b)]. Surprisingly, a systematically suppressed trajectory-to-trajectory standard deviation is observed when increasing the system size. The expectation value of the string operator, defined as $O_{\tn{str}}\equiv\prod_{i=2}^{L-1}(\sigma_{i-1}^x\sigma_i^z\sigma_{i+1}^x)$, also follows the exact reference result [Fig.~\ref{Fig:shadow_result}(c)] with a similarly controlled deviation and suppressed trajectory fluctuations [Fig.~\ref{Fig:shadow_result}(d)]. We have observed similar behavior for the MFIM for the energy and the longitudinal correlations observables, as shown in Fig.~\ref{Fig:shadow_result}(e--h) respectively. These benchmarks distinguish systematic error from fluctuations around the trajectory mean, motivating the analysis of trajectory typicality below.

\section{Bounds on error and trajectory typicality}

Consider the deviation $Z_\beta\equiv\rho_{\psi_\beta}-\rho_{\phi_\beta}$ between the sampled pure state $\rho_{\psi_\beta}\equiv|\psi_\beta\rangle\langle\psi_\beta|$ and the TFD reference state $\rho_{\phi_\beta}\equiv|\phi_\beta\rangle\langle\phi_\beta|$ at inverse temperature $\beta$, which separates into systematic and statistical errors as 
\be
\label{eq:deviation_decomp}
Z_\beta=Y_\beta+B_\beta;~~~
Y_\beta\equiv\rho_{\psi_\beta}-\mathbb{E}\rho_{\psi_\beta};~~~
B_\beta\equiv\mathbb{E}\rho_{\psi_\beta}-\rho_{\phi_\beta}.
\ee
Here $Y_\beta$ is the centered fluctuation relative to the averaged density matrix and $B_\beta$ is the deviation of the averaged density matrix relative from the reference. 
The late-time behavior of both $B_\beta$ and $Y_\beta$ follows from an exact formal solution of $Z_\beta$ (see Appendix), with randomness as source terms propagated by parallel transport~\cite{si}.

\begin{figure}[htb]
\centering
\includegraphics[width=\columnwidth]{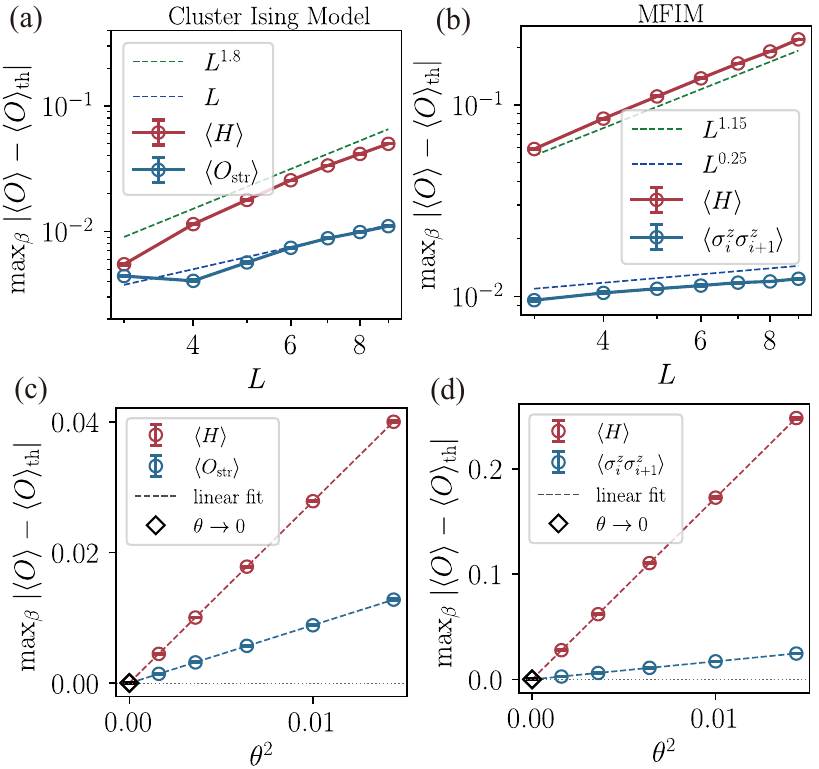}
\caption{\label{Fig:shadow_bias}\textbf{Scaling of systematic error} (a)-(b), with $L$, and (c)-(d), with $\theta$. Maximum absolute deviation of the trajectory-averaged observable from the exact thermal reference over the simulated temperature range, for \textbf{(a)}, the CIM, and \textbf{(b)}, the MFIM. Parameters are the same as Fig.~\ref{Fig:shadow_result}. Dashed lines are a guide to the eye.
Extrapolation to $\theta\rightarrow0$ at $L=5$ for 
\textbf{(c)}, the CIM and \textbf{(d)}, the MFIM. Both the scaling with $L$ and $\theta$ are consistent with Eq.~\eqref{eq:observable_bias_bound}.
}
\end{figure}

Let us start with the systematic error, $B_\beta = \mathbb{E}Z_\beta$, which includes~\cite{si} the truncation error in the Euler method, the group-commutator remainder, and the non-commutative error due to the non-linear evolution.
The error for the observable estimation, defined as $b_O(\beta)\equiv\Tr(O B_\beta)$, can be bounded exactly by the operator norm of the observable and the Hamiltonian.
We prove that the error in observable estimation is bounded~\cite{si,WangChowdhuryShadowGeometry2026} as,
\be
\label{eq:observable_bias_bound}
\big|b_O(\beta)\big|
\lesssim
\beta
\|\widetilde O\|_{\rm op}\|h\|_{\rm op}\big(|\theta|+|\alpha|\|h\|_{\rm op}\big)^2
\sim O\big(\theta^2\beta L^2\big),
\ee
where $\widetilde O=O-\langle \phi_\beta | O |\phi_\beta \rangle$, and we have assumed $|\theta|\simeq |\alpha|\|h\|_{\tn{op}}$, with $||h||_{\tn{op}}\sim L, ||\widetilde O||_{\tn{op}} \lesssim L$, i.e. $O$ is assumed to only have support on the system register. Remarkably, the error is controlled polynomial in both $\beta$ and $L$.
In Fig.~\ref{Fig:shadow_bias}a and \ref{Fig:shadow_bias}b, we show the maximum error relative to the true thermal expectation value over the simulated temperature range for both benchmark models, consistent with the predicted polynomial scaling with $L$.
In practice, by fixing $\theta=\alpha L$ and performing extrapolation to $\alpha,\theta\rightarrow 0$, the observable estimation is \textit{unbiased} (see Fig.~\ref{Fig:shadow_bias}c and \ref{Fig:shadow_bias}d), which highlights the \textit{exact} nature of the protocol upon extrapolation.

\begin{figure}[htb]
\centering
\includegraphics[width=0.95\columnwidth]{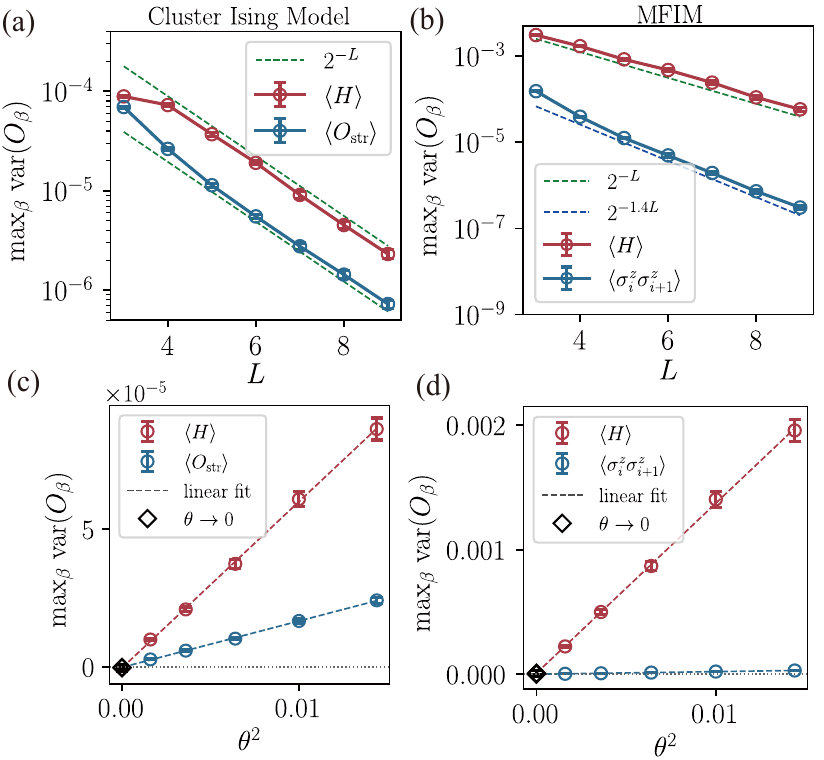}
\caption{\label{Fig:shadow_var}
\textbf{Scaling of trajectory variance} (a)-(b), with $L$, and (c)-(d), with $\theta$. Maximum trajectory variance over the simulated temperature range, for {\bf (a)}, the CIM, and {\bf (b)}, the MFIM. Parameters are the same as Fig.~\ref{Fig:shadow_result}. Dashed lines are a guide to the eye.
Extrapolation to $\theta\rightarrow0$ at $L=5$ for 
\textbf{(c)}, the CIM and \textbf{(d)}, the MFIM. Both the scaling with $L$ and $\theta$ are consistent with Eq.~\eqref{eq:traj_var_obs}.}
\end{figure}

Next, we turn to the centered fluctuation of the observable estimator among trajectories.
The centered observable fluctuations have trajectory-to-trajectory variance $y_O(\beta)\equiv\operatorname{Var}[\operatorname{Tr}(OY_\beta)]$. We prove~\cite{si,WangChowdhuryShadowGeometry2026} that 
\be
\label{eq:traj_var_obs}
y_O(\beta)
\lesssim
\frac{\alpha\theta \beta}{N} ||h||_{\tn{op}}^2 ||O||_{\tn{op}}^2
\sim O\bigg(\frac{\theta^2\beta L^3}{2^L}\bigg),
\ee
where $N=2^L$ is the Hilbert space dimension of the system register. Remarkably, the trajectory variance is exponentially suppressed in $L$.
A typical chain for observable estimation is therefore exponentially close to the trajectory average, which we refer to as the {\it trajectory typicality}.
As a corollary, the fluctuation in the squared Fubini-Study distance satisfies the same bounds on the trajectory variance, with an $O(1)$ prefactor.
In Fig.~\ref{Fig:shadow_var}, we show the maximum trajectory variance over the simulated temperature range for both benchmark models, consistent with the predicted exponential suppression with $L$ and the scaling with $\theta$.
Both Eq.~\eqref{eq:observable_bias_bound} and Eq.~\eqref{eq:traj_var_obs} are rigorously certified assuming $\theta\beta\|h\|_{\tn{op}}\sim O(1)$; in the benchmarks, the Trotter step-size $\theta$ is larger than the certified regime, whereas the same scalings are observed empirically.

\section{Intrinsic auxiliary-field path integral}

The essence of the shadow chain, namely one {\it single} shadow record per walker being sufficient to steer the evolution, originates from its fundamental connection to an auxiliary-field Euclidean path integral (see Appendix).
The collection of all the shadow histories before step $k$, denoted as $\Omega_k \equiv \{\omega_0, \omega_1, ..., \omega_{k-1}\}$, defines a stochastic trajectory on the pure state manifold whose probability distribution is given by
\begin{subequations}
	\label{eq:markov}
	\beq
    \label{eq:path_measure}
	&&p(\Omega_k)
	=
	p(\Omega_0)
	\prod_{j=1}^{k-1} p(b_j|U_j,\omega_{j-1},\psi_{j-1}) \mu(U_j),
	\\
	&&p(b_j|U_j,\omega_{j-1},\psi_{j-1})
	\equiv
	\big|\langle b_j| U_j^\dagger e^{\Delta\tau[P_{j-1},h]} |\psi_{j-1}\rangle\big|^2,~~~~~
	\eeq
\end{subequations}
with $\mu$ denoting the Haar measure, $p(\Omega_0)$ the initial distribution, and $\omega_j \equiv \{ U_j, \ket{b_j} \}$ the shadow record at step $j$. 
Since $\ket{\psi_j(\Omega_j)}$ enters Eq.~\eqref{eq:markov} only through the state itself, the shadow history factorizes and thus defines a Markovian process on the pure state manifold. 
Each shadow trajectory thus corresponds to a random walker whose increments are drawn by the intrinsic Born rule. Different random walkers are independent and their dynamics can be parallelized on different quantum processing units.

The probability distribution in Eq.~\eqref{eq:markov} defines an intrinsic Born-weighted path measure of shadow trajectories $\Omega_k$ on the pure-state manifold.
Averaging over many such trajectories defines a path-integral as,
\be
\label{eq:path_int}
\overline{\ket{\psi_{k}}} = \int \prod_{j=0}^{k-1} d\omega_j~ p(\Omega_k)  e^{\Delta\tau[P_{k-1|\omega_{k-1}},h]}\cdots e^{\Delta\tau[P_{0|\omega_{0}},h]} \ket{\psi_0},
\ee
where the action of $e^{\Delta\tau[P_{j|\omega_j},h]}$ and the averaging in $\overline{\ket{\psi_k}}$ conceptually denote the evolution and the trajectory-average of the corresponding pure-state density matrix, respectively. Here, $P_{k|\omega_k}$ is the shadow projector defined in Eq.~\eqref{eq:shadow_def_p} and is a function of the shadow record $\omega_k$.
In the language of determinant QMC,
the random shadow snapshot $P_k$ serves as the {\it auxiliary field}, the single-step shadow evolution replaces the Hubbard-Stratonovich (HS) transformation, and the path measure $p(\Omega_k)$ plays the role of the determinant weight of a Monte-Carlo configuration.
In sharp contrast, here the auxiliary variable $P_k$ is generated intrinsically by measurement without HS decoupling by hand. The path weight never needs to be explicitly evaluated; the Born rule samples each auxiliary variable directly, and thus the quantum state generates its own path measure on the fly.

\section{Resource scaling and extensions}

The exponential renormalization of the effective Trotter step size $
\Delta\tau_{\tn{eff}}
=\alpha\theta/(d+1)
$ is a direct consequence of dropping the shadow reconstruction in Eq.~\eqref{eq:path_int}, since it is not a physical completely-positive trace-preserving operation. 
Reaching a target inverse temperature $\beta$ requires $N_T=\beta(d+1)/(2\alpha\theta)$ steps; the qubits are required to remain coherent within the $N_T$ steps.
Choosing $\alpha^{-1}$ and $\theta^{-1}$ polynomial in $\beta$ and $L$ to control the error therefore keeps the step count $O(\tn{poly}(\beta, L)4^L)$.
Since each shadow step consumes a freshly-prepared state, reaching the step-$N_T$ requires a total $O(N_T^2)$ cooling steps per trajectory; replaying the full shadow history to collect all intermediate temperatures also requires $O(N_T^2)$ steps.  
For each shadow step, $e^{i\theta P} = U e^{i\theta|b\rangle\langle b|} U^{\dagger}$ compiles directly since the projector $P$ is classically known, Hamiltonian evolution can be implemented using $O(\operatorname{poly}(L))$ gates, whereas the exact global Haar random unitary $U_k$ uses $\Theta(16^{L})$ gates~\cite{ShendeBullockMarkov2006Synthesis}. Thus, estimating observables at all intermediate temperatures altogether requires $O(\operatorname{poly}(\beta,L)256^L)$ gates per random walker. Trajectory typicality allows the number of random walkers to be $O(1)$ asymptotically at large system size, so per-walker gate scaling {\it is} the total resource scaling. We have also proved rigorously that the Haar random unitary can be replaced by the Clifford 3-design~\cite{Zhu2017Clifford3Design,Webb2016Clifford3Design,KuengGross2015Stabilizer3Design} while keeping a polynomially controlled error and an exponentially suppressed variance, which reduces the per-shadow-step cost to $O(\tn{poly}(L))$~\cite{AaronsonGottesman2004StabilizerCircuits} and $O(\tn{poly}(\beta, L) 16^L)$ gates in total.

The observable estimations discussed above are obtained by replaying the full protocol and direct measurements. Alternatively, the shadow record can also be used to estimate many observables all-at-once. For multiple observables, the logarithmic dependence of sample complexity on their number carries over via the median-of-means (MoM) estimator. The stored snapshots $\omega_k$ from {\it different} trajectories can be combined to feed into the MoM estimator.
For each observable $O$, the total variance $v_O(k)$ relevant for the MoM estimator at step $k$ combines trajectory fluctuations $y_O(k)$ and the intrinsic Haar variance $v_O^{\rm sh}(\rho_{\psi_k})$ conditioned on a fixed trajectory as
$v_O(k)\equiv
y_O(k)
+\mathbb E_{\Omega_k}
\left[v_O^{\tn{sh}}(\rho_{\psi_k})\right]$\cite{si}.
For estimating $M$ observables, let $V_*(k)=\max_{1\leq m\leq M}v_{O_m}(k)$. The MoM estimator requires $O(V_*(k)\log M)$ independent trajectories at fixed accuracy and confidence.
Since trajectory variance is exponentially suppressed at a large system size $L$, the total variance is dominated by $v_O^{\tn{sh}}(\rho_{\psi_k})$, which can be exponentially large in $L$~\cite{HuangKuengPreskill2020}.

\section{Discussion}

We have established the shadow chain as an ansatz-free framework for quantum imaginary-time evolution with intrinsic stochasticity and controlled performance guarantees. A single shadow snapshot replaces in-situ estimation of state-dependent control parameters, while its classical record makes the trajectory repeatable. The Markovian dynamics realizes an intrinsic auxiliary-field path integral on quantum hardware, with the classical shadows serving as auxiliary fields. Thermal-observable estimates have controllable systematic error and exponentially suppressed trajectory fluctuations.

The advantage of the shadow chain is twofold. First, in the QMC language, trajectory typicality allows a single configuration to approach the ensemble prediction as system size grows; due to the Born-rule stochasticity, the shadow chain is intrinsically sign-problem-free for generic Hamiltonians. The $O(\tn{poly}(\beta, L) 16^L)$ gate-cost scaling confines the exponential cost only in $L$, which is more favorable at low temperature than the $e^{O({\beta L})}$ scaling of sign-problem reweighting\cite{Loh1990,TroyerWiese2005} of QMC.
Second, the shadow chain is ansatz-free, and its gate-cost scaling significantly improves performance at low temperature compared to recursive-compiled approaches.

The shadow-chain dynamics developed here also shares a similarity with field-theoretical response theory \cite{si}.
Ignoring Trotter errors and in an idealized continuous limit, with randomness injected along trajectories as source terms on a disordered background~\cite{WangChowdhuryShadowGeometry2026}. If the underlying field theory exists, our rigorous bound corresponds to the exact fixed-point, whereas a tadpole one-loop approximation around the free fixed-point exhibits the same scaling for both bounds in the error and the trajectory variance (see a companion article~\cite{WangChowdhuryShadowGeometry2026}). This motivates the conjecture of an intrinsic large-$N$ field theory governing the shadow chain dynamics for future work.

The shadow chain may extend beyond cooling, which is gradient descending Hamiltonian flow.
The ubiquity of gradient descent raises the question of whether shadow-chain-inspired stochastic dynamics can address other generic optimization problems in an {\it optimization-free} fashion. Potential applications include the quantum approximate optimization algorithm~\cite{FarhiGoldstoneGutmann2014}, quantum machine learning~\cite{Biamonte2017QML,Huang2025GenerativeQuantumAdvantage,LloydWeedbrook2018,Cong2019QCNN,Gao2022GenerativeCorrelations}, and, given the connection to Markovian diffusion on the pure-state manifold, quantum diffusion models for generative quantum-state learning~\cite{Zhang2024QuantumDiffusion,Liu2025MeasurementBasedQuantumDiffusion}.
We leave their exploration and the classification of problems admitting such implementations to future work.
\color{black}

{\bf Acknowledgments.}
XW thanks Haoyu Guo and Zhou Yang for bringing \cite{grover_qite} to his attention, which also partly inspired this work. XW acknowledges the use of large language models (ChatGPT 5.6 Sol and Claude Opus 4.6) for research assistance. XW and DC are supported in part by a grant from the Department of Energy (DE-SC0026112) under the Early Career Research Program.

{\bf Author contributions.}
XW conceived the project, designed the algorithm, and performed the theoretical and numerical studies. DC supervised the project, proposed the benchmarks, and provided insight into classical shadows during early discussions which partly inspired this work. XW and DC analyzed the data and wrote the manuscript.


{\bf Data and materials availability.}
 The data and code for this study have been deposited in the Zenodo database, available at \url{https://doi.org/10.5281/zenodo.23683180}.

\begin{appendices}
\appendix
\section{Classical shadow}
\label{meth:protocol}
Classical shadows encode information about a quantum state in classical records of randomized measurements~\cite{HuangKuengPreskill2020,Elben2023Toolbox}. For a density matrix $\rho$ on a $d$-dimensional register, we sample a Haar-random unitary $U$, apply $U^\dagger$, and measure the computational-basis outcome $b$. The record defines the projector $P=U\ket b\bra b U^\dagger$. Averaging over the unitary and the Born-rule outcomes gives the measurement channel defined as,
\be
\mathcal M(\rho) \equiv \mathbb{E}[P] = \int_{\mu(U)} \sum_b \tn{Tr}(P \rho) P = \frac{\rho + \mathbb{1}}{d+1},
\ee
where $\mu(U)$ is the corresponding random unitary ensemble.
The channel $\mathcal M(\rho)$ is non-invertible, and conceptually the shadow estimator $\widehat\rho=\mathcal M^{-1}(P)=(d+1)P-\mathbb 1$ is achieved by classical post-processing.

The higher Haar moments are evaluated using Weingarten calculus~\cite{CollinsSniady2006}, which expresses integrals of products of unitary matrix elements as sums over permutations. The Born probability contributes an additional factor of $P$, so the Born-weighted $q-$th moment of $P$ requires the $(q+1)-$th moment of a uniformly sampled Haar projector. For any integer $q\geq1$, we obtain
\beq
\label{eq:methods_shadow_qth_moment}
\mathbb E\!\left[P^{\otimes q}\mid\rho\right]
=\frac{\sum_{\pi\in S_{q+1}}
\operatorname{Tr}_0\!\left[(\rho\otimes\mathbb 1^{\otimes q})U_\pi\right]}
{\prod_{j=1}^{q}(d+j)}.
\eeq
Here $S_{q+1}$ is the permutation group on the $q+1$ tensor factors, $U_\pi$ implements the permutation $\pi$, and $\operatorname{Tr}_0$ traces out the first factor carrying $\rho$.
Global Haar-random is exact, while we prove \cite{si} that Clifford 3-design is enough for retaining all the results above.

\section{Double-bracket imaginary-time evolution}
\label{meth:ite}
The imaginary-time evolution under $h$ for a density matrix $\rho$ is a normalized flow defined as,
\begin{equation}
\label{eq:ite_map}
\mathcal T_{\tau+\Delta\tau,\tau}[\rho]
=\frac{e^{-\Delta\tau h}\rho e^{-\Delta\tau h}}
{\Tr(e^{-2\Delta\tau h}\rho)}.
\end{equation}
Under assumptions of purity, the flow is equivalent \cite{si} to a double bracket ITE equation for the density matrix \cite{Gluza2026DBQITE}.
For a pure state $\rho_\phi=\ket{\phi}\bra{\phi}$, this flow satisfies the differential equation $\partial_\tau\rho_\phi=[[\rho_\phi,h],\rho_\phi]$. Freezing this generator over a step of size $\Delta\tau=\alpha\theta$ gives a state-dependent unitary,
\be
\label{eq:db_ite_trotter}
\ket{\phi_{k+1}} = e^{\Delta\tau[\rho_{\phi_k},h]}\ket{\phi_k};
\qquad \rho_{\phi_k}\equiv\ket{\phi_k}\bra{\phi_k},
\ee
which is a Riemannian gradient descent of $\Tr(h\rho)$ on the pure-state
manifold~\cite{Gluza2026DBQITE,McMahonPervezArenz2026}. A gate-level realization of this unitary is the four-pulse group commutator
\be
\label{eq:shadow_evolve_approx}
e^{\Delta\tau[\rho_{\phi_k}, h]} =
e^{i\alpha h}e^{i\theta\rho_{\phi_k}}e^{-i\alpha h}e^{-i\theta\rho_{\phi_k}}
+ O(\alpha^2\theta,\alpha\theta^2).
\ee
A higher order group commutator can be also used to reduce the Trotter error~\cite{ChildsWiebe2013Commutators}, which is beyond the scope of this work.

\section{Auxiliary-shadow ITE}
\label{meth:auxiliary_shadow}
The discretized Hubbard stratonovich (HS) transformation is a standard tool in QMC to decouple the two-body interaction term in the Hamiltonian into a one-body term coupled to an auxiliary field. Let us denote the interaction of interest scehmatically as a perfect square, $H_{\tn{int}}=g\Gamma^2$, where $\Gamma$ is a one-body fermion-bilinear operator and $g$ is the coupling constant. The HS transformation of a single ITE step is then given by
\be
\label{eq:hs}
e^{-\Delta\tau H_{\tn{int}}} \simeq \sum_{\Phi} \gamma(\Phi) e^{\sqrt{-\Delta\tau g}\Phi \Gamma},
\ee
where $\Phi$ is the auxiliary field and $\gamma(\Phi)$ is the corresponding weight. The sum over the auxiliary field $\Phi$ is typically performed stochastically by Monte Carlo sampling.

The single-step shadow evolution in Eq.~\eqref{eq:shadow_evolve} shares a similar structure, defined as
\be
\label{eq:aux_shadow_ite}
e^{\Delta\tau_{\tn{eff}}[\rho_{\psi_{\tau}},h]}\ket{\psi_{\tau}} 
\simeq
 \int_{\mu(U)} \sum_b \tn{Tr}(P \rho_{\psi_\tau}) e^{\Delta\tau[P,h]}\ket{\psi_{\tau}},
\ee
where $P$ is the shadow projector defined in Eq.~\eqref{eq:shadow_def_p} and $\Delta\tau_{\tn{eff}} = \Delta\tau/(d+1)$ is the effective time step. The shadow projector $P$ serves as an auxiliary field, and the Born probability $\tn{Tr}(P \rho_{\psi_\tau})$ serves as the corresponding weight. The sum over the auxiliary shadow $P$ is performed intrinsically by the Born rule, without any need for Monte Carlo sampling. Therefore, the auxiliary-shadow ITE in Eq.~\eqref{eq:aux_shadow_ite} connects closely to the HS-decomposed stochastic ITE in QMC.

\section{Euclidean path integral}
\label{meth:path_integral}

The auxiliary-field interpretation of the shadow evolution motivates us to formulate a Euclidean path integral on the pure-state manifold. A concrete expression is shown in Eq.~\eqref{eq:path_int}, and we provide a separate point of view via inserting the resolution of identity  here.

Denoting the shadow channel as $\mathcal{M}[\rho]\equiv \mathbb{E}_{U,b}[P]$ and its conceptual inverse as $\mathcal{M}^{-1}[\rho]$, the resolution of identity $\mathbb{1}_{\mathcal{M}} = \mathcal{M}^{-1} \mathcal{M}$ cast the shadow evolution form that is conceptually equivalent to the imaginary-time path integral,
\be
\label{eq:pi_ite}
\ket{\psi_{k}}
=
e^{-\Delta\tau H}
\mathbb{1}_{\mathcal{M}}
\cdots 
\mathbb{1}_{\mathcal{M}}
e^{-\Delta\tau H}
\mathbb{1}_{\mathcal{M}}
e^{-\Delta\tau H} \ket{\psi_0}.
\ee
Here, the action of $\mathbb{1}_{\mathcal{M}}$ and its inverse on a ket denote the action on its pure state density matrix.
Since the shadow channel $\mathcal{M}$ is non-invertible, its inverse $\mathcal{M}^{-1}$ cannot be achieved coherently and are thus dropped in the protocol. This non-invertibility translates to the exponentially renormalized effective Trotter step-size $
\Delta\tau_{\tn{eff}}
=\alpha\theta/(d+1)
$, sharing the same origin as the signal-to-noise issue in the tomographical classical shadow.

\section{Martingale and stochastic differential equation}
\label{meth:sde}
Let us now formulate the shadow chain as a stochastic differential equation (SDE) with a generic martingale. At step $k$, the Doob decomposition~\cite{Williams1991} separates the conditionally averaged flow from a fresh stochastic increment, and gives the SDE as,
\begin{subequations}
\beq
\label{eq:ket_sde}
\ket{\psi_{k+1}} 
&=& 
\ket{\psi_{k}} + \eta[\rho_{\psi_k}, h]\ket{\psi_k} + \ket{\xi_{k+1}},
\\
\ket{\xi_{k+1}} &=& \alpha\theta \bigg[P_{k} - \frac{\rho_{\psi_k}+\mathbb{1}}{d+1}, h\bigg]\ket{\psi_k},
\eeq
\end{subequations}
where $\rho_{\psi_k}=\ket{\psi_k}\bra{\psi_k}$, $d=4^L$ is the dimension on which the global shadow is taken, and $\eta=\alpha\theta/(d+1)$. By construction, $\mathbb{E}[\ket{\xi_{k+1}}|\Omega_k]=0$ and thus is a martingale difference. 

Let us define a imaginary time grid $\tau_k=k\eta$, 
and {\it continuously} evolve (rather than trotterized evolve) between the discrete time steps using Eq.~\eqref{eq:ite_map} to obtain the exact TFD reference state $\ket{\phi_k}$.	
It is convenient to promote the SDE in Eq.~\eqref{eq:ket_sde} to generic density matrix form, and model it as driven parallel transport as,
\be
\label{eq:sde_mat_z_a}
\rho_{\psi_{k+1}}
=
\mathcal T_{k+1,k}[\rho_{\psi_k}]
+J_{k+1};\qquad
J_k \equiv
\big[\Xi_{k}^{(\psi)} + \varepsilon_{k}\big],
\ee
where $J_k$ is the source term that collect the matrix martingale $\Xi_{k+1}^{(\psi)}\equiv\ket{\xi_k}\bra{\psi_k}+\tn{h.c.}$ and the gate implementation error $\epsilon_k$, including the Euler error and the Baker–Campbell–Hausdorff (BCH) error. Denoting the deviation between the two pure state density matrices as
$
Z_k = \rho_{\psi_k} - \rho_{\phi_k},
$
an induced SDE can be formulated for the sample deviation $Z_k$ as
\begin{subequations}
\beq
\label{eq:cont_sol_z}
Z_{\tau+d\tau}
&=&
D_{\tau+d\tau,\tau}[Z_\tau] + J_{\tau+d\tau}
\\
D_{\tau+d\tau,\tau}[Z_\tau]
&\simeq&
Z_\tau + \mathcal{L}_\tau[Z_\tau] d\tau + F(Z_\tau)d\tau 
\\
\mathcal{L}_\tau[Z_\tau]
&\equiv&
[[Z_\tau,h],\rho_{\phi_\tau}]+[[\rho_{\phi_\tau},h],Z_\tau]~~~~
\\
F(Z_\tau)&\equiv&[[Z_\tau,h],Z_\tau]
\eeq
\end{subequations}
The deterministic part of the SDE has a concrete physical meaning: given a continuous family of deterministic flows, $D_{f,i}[Z_i]$ conceptually describes how a separation $Z_i$ connecting two nearby deterministic flow trajectories at step $i$ evolves into their new separation at step $f$. There are two contributions to this evolution of $Z_\tau$: the linear term $\mathcal{L}_\tau[Z_\tau]$ describes the parallel transport of the deviation along the reference trajectory, while the nonlinear term $F(Z_\tau)$ describes the curvature of the state space. A post-kick driving term $J_{\tau+d\tau}$ is injected at each step.

It can be shown that the nonlinear term is the source of a controllable systematic error in observable estimation\cite{si}, specifically due to the non-commutativity between the expectation value and the nonlinear evolution, i.e. $[\mathbb{E}, F]\neq 0$. 

\section{Exact field-theoretical solution}
\label{meth:exact_solution}

Remarkably, the deterministic part of Eq.~\eqref{eq:sde_mat_z_a} can be solved exactly by the late-time push-forward $D_{f,i}[Z_i]$ as a resolvent.
For $f\geq i$, the normalized propagator of the reference flow is defined as,
\begin{equation}
\label{eq:methods_propagator}
G_{f,i}^{\phi}=
\frac{e^{-(\tau_f-\tau_i)h}}
{\|e^{-(\tau_f-\tau_i)h}\ket{\phi_{i}}\|},
\end{equation}
where $\tau_i$ is the $i$-th point on the imaginary-time grid. 
The exact change of a traceless perturbation under normalized propagation is
\begin{equation}
\label{eq:methods_pushforward}
D_{f,i}[Z]=
\frac{G_{f,i}^{\phi}ZG_{f,i}^{\phi}
-\rho_{\phi_f}\Tr(G_{f,i}^{\phi}ZG_{f,i}^{\phi})}
{1+\Tr(G_{f,i}^{\phi}ZG_{f,i}^{\phi})},
\end{equation}
which enters Eq.~\eqref{eq:cont_sol_z}. The denominator retains the normalization nonlinearity exactly. As a resolvent, Eq.~\eqref{eq:methods_pushforward} features a  M\"obius-like property and thereby allows us to cast this formal exact solution~\cite{si} to the SDE defined in Eq.~\eqref{eq:sde_mat_z_a} as
\begin{subequations}
\begin{align}
\label{eq:sol_x_tot}
X_k&=\sum_{1\leq i<k}\big[1+\Tr(X_i)\big]
G_{k,i}^{\phi}J_iG_{k,i}^{\phi},\\
\label{eq:sol_z_tot}
Z_k&=\frac{X_k-\rho_{\phi_k}\Tr(X_k)}{1+\Tr(X_k)}+J_k,
\end{align}
\end{subequations}
assuming the denominators are nonzero and with the initial condition $X_0=Z_0=J_0=0$. This representation propagates and accumulates all earlier sources to the final step and retains the fresh source at the final step. It is the starting point for bounding the observable moments.

\par
\end{appendices}
\counterwithout*{equation}{section}

\newpage

\bibliographystyle{apsrev4-1_custom}
\bibliography{refs}

@misc{grover_qite,
      title={Grover's algorithm is an approximation of imaginary-time evolution}, 
      author={Yudai Suzuki and Marek Gluza and Jeongrak Son and Bi Hong Tiang and Nelly H. Y. Ng and Zoë Holmes},
      year={2026},
      eprint={2507.15065},
      archivePrefix={arXiv},
      primaryClass={quant-ph},
      url={https://arxiv.org/abs/2507.15065}, 
}

@article{TroyerWiese2005,
  author  = {Troyer, Matthias and Wiese, Uwe-Jens},
  title   = {Computational Complexity and Fundamental Limitations to Fermionic Quantum {M}onte {C}arlo Simulations},
  journal = {Phys. Rev. Lett.},
  volume  = {94},
  pages   = {170201},
  year    = {2005},
  doi     = {10.1103/PhysRevLett.94.170201}
}

@incollection{AssaadEvertz2008,
  author    = {Assaad, F. F. and Evertz, H. G.},
  title     = {{World-line and Determinantal Quantum Monte Carlo Methods for Spins, Phonons and Electrons}},
  booktitle = {Computational Many-Particle Physics},
  series    = {Lecture Notes in Physics},
  volume    = {739},
  editor    = {Fehske, H. and Schneider, R. and Wei{\ss}e, A.},
  publisher = {Springer},
  address   = {Berlin, Heidelberg},
  pages     = {277--356},
  year      = {2008},
  doi       = {10.1007/978-3-540-74686-7_10}
}

@article{ZhangKrakauer2003,
  author  = {Zhang, Shiwei and Krakauer, Henry},
  title   = {Quantum {M}onte {C}arlo Method using Phase-Free Random Walks with {S}later Determinants},
  journal = {Phys. Rev. Lett.},
  volume  = {90},
  pages   = {136401},
  year    = {2003},
  doi     = {10.1103/PhysRevLett.90.136401}
}

@inproceedings{GilyenSuLowWiebe2019QSVT,
  author        = {Gily{\'e}n, Andr{\'a}s and Su, Yuan and Low, Guang Hao and Wiebe, Nathan},
  title         = {Quantum singular value transformation and beyond: exponential improvements for quantum matrix arithmetics},
  booktitle     = {Proc. 51st Annual ACM Symposium on Theory of Computing (STOC 2019)},
  pages         = {193--204},
  year          = {2019},
  doi           = {10.1145/3313276.3316366},
  eprint        = {1806.01838},
  archivePrefix = {arXiv}
}

@article{Motta2020QITE,
  title   = {Determining eigenstates and thermal states on a quantum computer using quantum imaginary time evolution},
  author  = {Motta, Mario and Sun, Chong and Tan, Adrian T. K. and O'Rourke, Matthew J. and Ye, Erika and Minnich, Austin J. and Brand{\~a}o, Fernando G. S. L. and Chan, Garnet Kin-Lic},
  journal = {Nature Physics},
  volume  = {16},
  number  = {2},
  pages   = {205--210},
  year    = {2020}
}

@article{McMahonPervezArenz2026,
  title = {Equating quantum imaginary time evolution, Riemannian gradient flows, and stochastic implementations},
  author = {McMahon, Nathan A. and Pervez, Mahum and Arenz, Christian},
  journal = {Phys. Rev. Res.},
  volume = {8},
  issue = {2},
  pages = {023024},
  numpages = {6},
  year = {2026},
  month = {Apr},
  publisher = {American Physical Society},
  doi = {10.1103/ht2m-1j91},
  url = {https://link.aps.org/doi/10.1103/ht2m-1j91}
}

@misc{McMahonPervezArenz_arxiv,
      title={Riemannian gradient descent-based quantum algorithms for ground state preparation with guarantees}, 
      author={Mahum Pervez and Ariq Haqq and Nathan A. McMahon and Christian Arenz},
      year={2025},
      eprint={2512.13401},
      archivePrefix={arXiv},
      primaryClass={quant-ph},
      url={https://arxiv.org/abs/2512.13401}, 
}

@misc{dq2mc,
      title={Determinant Quantum-Quantum Monte Carlo: Coherent Auxiliary-Field Sampling}, 
      author={Xuepeng Wang and Sagnik Banerjee and Debanjan Chowdhury},
      year={2026},
      eprint={2608.28742},
      archivePrefix={arXiv},
      primaryClass={cond-mat.str-el},
      url={https://arxiv.org/abs/2608.28742}, 
}

@misc{si, title={See supplementary material for additional details.}}

@article{Gluza2026DBQITE,
  title   = {Double-Bracket Quantum Algorithms for Quantum Imaginary-Time Evolution},
  author  = {Gluza, Marek and Son, Jeongrak and Tiang, Bi Hong and Zander, Ren{\'e} and Seidel, Raphael and Suzuki, Yudai and Holmes, Zo{\"e} and Ng, Nelly H. Y.},
  journal = {Phys. Rev. Lett.},
  volume  = {136},
  pages   = {020601},
  year    = {2026},
  doi     = {10.1103/rw81-k8vk}
}

@article{Brockett1991DoubleBracket,
  author  = {Brockett, R. W.},
  title   = {Dynamical systems that sort lists, diagonalize matrices, and solve linear programming problems},
  journal = {Linear Algebra Appl.},
  volume  = {146},
  pages   = {79--91},
  year    = {1991},
  doi     = {10.1016/0024-3795(91)90021-N}
}

@article{Wegner1994FlowEquations,
  author  = {Wegner, Franz},
  title   = {Flow-equations for {Hamiltonians}},
  journal = {Ann. Phys. (Berlin)},
  volume  = {3},
  number  = {2},
  pages   = {77--91},
  year    = {1994},
  doi     = {10.1002/andp.19945060203}
}

@article{Gluza2024Diagonalization,
  author  = {Gluza, Marek},
  title   = {Double-bracket quantum algorithms for diagonalization},
  journal = {Quantum},
  volume  = {8},
  pages   = {1316},
  year    = {2024},
  doi     = {10.22331/q-2024-04-09-1316}
}

@article{Kolotouros2025RMITE,
  author  = {Kolotouros, Ioannis and Joseph, David and Narayanan, Anand Kumar},
  title   = {Accelerating quantum imaginary-time evolution with random measurements},
  journal = {Phys. Rev. A},
  volume  = {111},
  number  = {1},
  pages   = {012424},
  year    = {2025},
  doi     = {10.1103/PhysRevA.111.012424}
}

@article{Suzuki2025DBQSP,
  author  = {Suzuki, Yudai and Tiang, Bi Hong and Son, Jeongrak and Ng, Nelly H. Y. and Holmes, Zoe and Gluza, Marek},
  title   = {Double-bracket algorithm for quantum signal processing without post-selection},
  journal = {Quantum},
  volume  = {9},
  pages   = {1954},
  year    = {2025},
  doi     = {10.22331/q-2025-12-23-1954}
}

@misc{KolotourosGarciaPatron2026PDBQITE,
  author        = {Kolotouros, Ioannis and Garcia-Patron, Raul},
  title         = {A quantum double-bracket algorithm for imaginary-time evolution with exponentially shorter depth},
  year          = {2026},
  eprint        = {2609.37095},
  archivePrefix = {arXiv},
  primaryClass  = {quant-ph},
  doi           = {10.48550/arXiv.2609.37095},
  url           = {https://arxiv.org/abs/2609.37095}
}

@article{BrodyHughston2001GeometricQM,
  title   = {Geometric quantum mechanics},
  author  = {Brody, Dorje C. and Hughston, Lane P.},
  journal = {J. Geom. Phys.},
  volume  = {38},
  pages   = {19--53},
  year    = {2001},
  doi     = {10.1016/S0393-0440(00)00052-8},
  eprint  = {quant-ph/9906086},
  archivePrefix = {arXiv}
}

@misc{Zander2025RiemannianDBQITE,
  title   = {Role of {Riemannian} geometry in double-bracket quantum imaginary-time evolution},
  author  = {Zander, Ren{\'e} and Seidel, Raphael and Li, Xiaoyue and Gluza, Marek},
  year    = {2025},
  eprint  = {2504.01065},
  archivePrefix = {arXiv},
  primaryClass = {quant-ph},
  url     = {https://arxiv.org/abs/2504.01065}
}

@article{Son2025QDP,
  title   = {Quantum Dynamic Programming},
  author  = {Son, Jeongrak and Gluza, Marek and Takagi, Ryuji and Ng, Nelly H. Y.},
  journal = {Phys. Rev. Lett.},
  volume  = {134},
  pages   = {180602},
  year    = {2025},
  doi     = {10.1103/PhysRevLett.134.180602}
}

@article{SugiuraShimizu2012TPQ,
  title   = {Thermal Pure Quantum States at Finite Temperature},
  author  = {Sugiura, Sho and Shimizu, Akira},
  journal = {Phys. Rev. Lett.},
  volume  = {108},
  pages   = {240401},
  year    = {2012}
}

@article{SugiuraShimizu2013TPQ,
  title   = {Canonical Thermal Pure Quantum State},
  author  = {Sugiura, Sho and Shimizu, Akira},
  journal = {Phys. Rev. Lett.},
  volume  = {111},
  pages   = {010401},
  year    = {2013}
}

@article{Coopmans2023PureThermalShadows,
  author  = {Coopmans, Luuk and Kikuchi, Yuta and Benedetti, Marcello},
  title   = {Predicting {Gibbs}-state expectation values with pure thermal shadows},
  journal = {PRX Quantum},
  volume  = {4},
  pages   = {010305},
  year    = {2023},
  doi     = {10.1103/PRXQuantum.4.010305}
}

@article{White2009METTS,
  author  = {White, Steven R.},
  title   = {Minimally entangled typical quantum states at finite temperature},
  journal = {Phys. Rev. Lett.},
  volume  = {102},
  pages   = {190601},
  year    = {2009},
  doi     = {10.1103/PhysRevLett.102.190601}
}

@article{StoudenmireWhite2010METTS,
  author  = {Stoudenmire, E. M. and White, Steven R.},
  title   = {Minimally entangled typical thermal state algorithms},
  journal = {New J. Phys.},
  volume  = {12},
  pages   = {055026},
  year    = {2010},
  doi     = {10.1088/1367-2630/12/5/055026}
}

@article{PoulinWocjan2009Gibbs,
  title   = {Sampling from the Thermal Quantum {G}ibbs State and Evaluating Partition Functions with a Quantum Computer},
  author  = {Poulin, David and Wocjan, Pawel},
  journal = {Phys. Rev. Lett.},
  volume  = {103},
  pages   = {220502},
  year    = {2009}
}

@article{Temme2011Metropolis,
  title   = {Quantum {M}etropolis sampling},
  author  = {Temme, K. and Osborne, T. J. and Vollbrecht, K. G. and Poulin, D. and Verstraete, F.},
  journal = {Nature},
  volume  = {471},
  pages   = {87--90},
  year    = {2011}
}

@article{YungAspuruGuzik2012Metropolis,
  author  = {Yung, Man-Hong and Aspuru-Guzik, Al{\'a}n},
  title   = {A quantum--quantum {Metropolis} algorithm},
  journal = {Proc. Natl. Acad. Sci. U.S.A.},
  volume  = {109},
  pages   = {754--759},
  year    = {2012},
  doi     = {10.1073/pnas.1111758109}
}

@article{ChowdhurySomma2017Gibbs,
  title   = {Quantum algorithms for {G}ibbs sampling and hitting-time estimation},
  author  = {Chowdhury, Anirban N. and Somma, Rolando D.},
  journal = {Quantum Information and Computation},
  volume  = {17},
  number  = {1\&2},
  pages   = {41--64},
  year    = {2017}
}

@article{BSS1981,
  title   = {{M}onte {C}arlo calculations of coupled boson-fermion systems. I},
  author  = {Blankenbecler, R. and Scalapino, D. J. and Sugar, R. L.},
  journal = {Phys. Rev. D},
  volume  = {24},
  pages   = {2278--2286},
  year    = {1981}
}

@article{Huggins2022NatureQCQMC,
  title   = {Unbiasing fermionic quantum {M}onte {C}arlo with a quantum computer},
  author  = {Huggins, William J. and O'Gorman, Bryan A. and Rubin, Nicholas C. and Reichman, David R. and Babbush, Ryan and Lee, Joonho},
  journal = {Nature},
  volume  = {603},
  pages   = {416--420},
  year    = {2022}
}

@article{Layden2023QuantumEnhancedMCMC,
  author  = {Layden, David and Mazzola, Guglielmo and Mishmash, Ryan V. and Motta, Mario and Wocjan, Pawel and Kim, Jin-Sung and Sheldon, Sarah},
  title   = {Quantum-enhanced {Markov} chain {Monte Carlo}},
  journal = {Nature},
  volume  = {619},
  pages   = {282--287},
  year    = {2023},
  doi     = {10.1038/s41586-023-06095-4}
}

@article{Wan2023MatchgateShadows,
  author  = {Wan, Kianna and Huggins, William J. and Lee, Joonho and Babbush, Ryan},
  title   = {Matchgate shadows for fermionic quantum simulation},
  journal = {Commun. Math. Phys.},
  volume  = {404},
  pages   = {629--700},
  year    = {2023},
  doi     = {10.1007/s00220-023-04844-0}
}

@article{Zhang2025QuantumComputingQMC,
  author        = {Zhang, Yukun and Huang, Yifei and Sun, Jinzhao and Lv, Dingshun and Yuan, Xiao},
  title         = {Quantum computing quantum {Monte Carlo} algorithm},
  journal       = {Phys. Rev. A},
  volume        = {112},
  pages         = {022428},
  year          = {2025},
  doi           = {10.1103/jt8s-hzhd},
  eprint        = {2206.10431},
  archivePrefix = {arXiv},
  primaryClass  = {quant-ph}
}

@misc{ChenKastoryanoGilyen2024GibbsSampler,
  title         = {An efficient and exact noncommutative quantum {G}ibbs sampler},
  author        = {Chen, Chi-Fang and Kastoryano, Michael J. and Gily{\'e}n, Andr{\'a}s},
  year          = {2023},
  eprint        = {2311.09207},
  archivePrefix = {arXiv},
  primaryClass  = {quant-ph},
  doi           = {10.48550/arXiv.2311.09207},
  url           = {https://arxiv.org/abs/2311.09207}
}

@misc{ChenKastoryanoBrandaoGilyen2023ThermalPrep,
  title         = {Quantum thermal state preparation},
  author        = {Chen, Chi-Fang and Kastoryano, Michael J. and Brand{\~a}o, Fernando G. S. L. and Gily{\'e}n, Andr{\'a}s},
  year          = {2023},
  eprint        = {2303.18224},
  archivePrefix = {arXiv},
  primaryClass  = {quant-ph},
  doi           = {10.48550/arXiv.2303.18224},
  url           = {https://arxiv.org/abs/2303.18224}
}

@article{DingLiLin2025KMSGibbsSampler,
  author        = {Ding, Zhiyan and Li, Bowen and Lin, Lin},
  title         = {Efficient quantum {Gibbs} samplers with {Kubo--Martin--Schwinger} detailed balance condition},
  journal       = {Commun. Math. Phys.},
  volume        = {406},
  pages         = {67},
  year          = {2025},
  doi           = {10.1007/s00220-025-05235-3},
  eprint        = {2404.05998},
  archivePrefix = {arXiv},
  primaryClass  = {quant-ph}
}

@misc{Bakshi2024HighTemperatureGibbs,
  author        = {Bakshi, Ainesh and Liu, Allen and Moitra, Ankur and Tang, Ewin},
  title         = {High-temperature {Gibbs} states are unentangled and efficiently preparable},
  year          = {2024},
  eprint        = {2403.16850},
  archivePrefix = {arXiv},
  primaryClass  = {quant-ph},
  doi           = {10.48550/arXiv.2403.16850},
  url           = {https://arxiv.org/abs/2403.16850}
}

@article{LinTong2020NearOptimal,
  title   = {Near-optimal ground state preparation},
  author  = {Lin, Lin and Tong, Yu},
  journal = {Quantum},
  volume  = {4},
  pages   = {372},
  year    = {2020}
}

@article{DongLinTong2022Eigenvalue,
  title   = {Ground-state preparation and energy estimation on early fault-tolerant quantum computers via quantum eigenvalue transformation of unitary matrices},
  author  = {Dong, Yulong and Lin, Lin and Tong, Yu},
  journal = {PRX Quantum},
  volume  = {3},
  pages   = {040305},
  year    = {2022}
}

@article{Feynman1948,
  author  = {Feynman, R. P.},
  title   = {Space-Time Approach to Non-Relativistic Quantum Mechanics},
  journal = {Rev. Mod. Phys.},
  volume  = {20},
  pages   = {367},
  year    = {1948},
  doi     = {10.1103/RevModPhys.20.367}
}

@article{Kac1949,
  author  = {Kac, M.},
  title   = {On Distributions of Certain {W}iener Functionals},
  journal = {Trans. Amer. Math. Soc.},
  volume  = {65},
  pages   = {1},
  year    = {1949},
  doi     = {10.1090/S0002-9947-1949-0027960-X}
}

@article{SandvikKurkijarvi1991,
  author  = {Sandvik, A. W. and Kurkij{\"a}rvi, J.},
  title   = {Quantum {M}onte {C}arlo simulation method for spin systems},
  journal = {Phys. Rev. B},
  volume  = {43},
  pages   = {5950},
  year    = {1991},
  doi     = {10.1103/PhysRevB.43.5950}
}

@article{Loh1990,
  author  = {Loh, E. Y. and Gubernatis, J. E. and Scalettar, R. T. and White, S. R. and Scalapino, D. J. and Sugar, R. L.},
  title   = {Sign problem in the numerical simulation of many-electron systems},
  journal = {Phys. Rev. B},
  volume  = {41},
  pages   = {9301},
  year    = {1990},
  doi     = {10.1103/PhysRevB.41.9301}
}

@article{McArdle2019VarQITE,
  author  = {McArdle, S. and Jones, T. and Endo, S. and Li, Y. and Benjamin, S. C. and Yuan, X.},
  title   = {Variational ansatz-based quantum simulation of imaginary time evolution},
  journal = {npj Quantum Inf.},
  volume  = {5},
  pages   = {75},
  year    = {2019},
  doi     = {10.1038/s41534-019-0187-2}
}

@article{Yuan2019Theory,
  author  = {Yuan, X. and Endo, S. and Zhao, Q. and Li, Y. and Benjamin, S. C.},
  title   = {Theory of variational quantum simulation},
  journal = {Quantum},
  volume  = {3},
  pages   = {191},
  year    = {2019},
  doi     = {10.22331/q-2019-10-07-191}
}

@article{Mao2023MeasurementITE,
  author  = {Mao, Yuping and Chaudhary, Manish and Kondappan, Manikandan and Shi, Junheng and Ilo-Okeke, Ebubechukwu O. and Ivannikov, Valentin and Byrnes, Tim},
  title   = {Measurement-based deterministic imaginary time evolution},
  journal = {Phys. Rev. Lett.},
  volume  = {131},
  number  = {11},
  pages   = {110602},
  year    = {2023},
  doi     = {10.1103/PhysRevLett.131.110602}
}

@article{Schwartzman2026MultiCopyITE,
  author  = {Schwartzman, Tal and Zache, Torsten V. and Pichler, Hannes and Sadeghpour, H. R.},
  title   = {Imaginary time evolution and ground state preparation using unitary multi-copy protocols},
  journal = {PRX Quantum},
  volume  = {7},
  number  = {3},
  pages   = {033033},
  year    = {2026},
  doi     = {10.1103/1tf6-bc55}
}

@article{HuoLi2023MonteCarloITE,
  author  = {Huo, Mingxia and Li, Ying},
  title   = {Error-resilient {Monte Carlo} quantum simulation of imaginary time},
  journal = {Quantum},
  volume  = {7},
  pages   = {916},
  year    = {2023},
  doi     = {10.22331/q-2023-02-09-916}
}

@article{Magann2022FALQON,
  author  = {Magann, Alicia B. and Rudinger, Kenneth M. and Grace, Matthew D. and Sarovar, Mohan},
  title   = {Feedback-based quantum optimization},
  journal = {Phys. Rev. Lett.},
  volume  = {129},
  pages   = {250502},
  year    = {2022},
  doi     = {10.1103/PhysRevLett.129.250502}
}

@article{Magann2023RandomizedAdaptive,
  author  = {Magann, Alicia B. and Economou, Sophia E. and Arenz, Christian},
  title   = {Randomized adaptive quantum state preparation},
  journal = {Phys. Rev. Research},
  volume  = {5},
  pages   = {033227},
  year    = {2023},
  doi     = {10.1103/PhysRevResearch.5.033227}
}

@article{WiersemaKilloran2023RiemannianFlow,
  author  = {Wiersema, Roeland and Killoran, Nathan},
  title   = {Optimizing quantum circuits with {Riemannian} gradient flow},
  journal = {Phys. Rev. A},
  volume  = {107},
  pages   = {062421},
  year    = {2023},
  doi     = {10.1103/PhysRevA.107.062421}
}

@misc{Arulandu2026TITE,
  author        = {Arulandu, Alvan and Martyn, John M. and Chuang, Isaac L.},
  title         = {Trading Imaginary Time for Randomness in Ground State Preparation},
  year          = {2026},
  eprint        = {2608.00443},
  archivePrefix = {arXiv},
  primaryClass  = {quant-ph},
  doi           = {10.48550/arXiv.2608.00443},
  url           = {https://arxiv.org/abs/2608.00443}
}

@article{Aaronson2020ShadowTomography,
  author        = {Aaronson, Scott},
  title         = {Shadow tomography of quantum states},
  journal       = {SIAM J. Comput.},
  volume        = {49},
  number        = {5},
  pages         = {STOC18-368--STOC18-394},
  year          = {2020},
  doi           = {10.1137/18M120275X},
  eprint        = {1711.01053},
  archivePrefix = {arXiv},
  primaryClass  = {quant-ph},
  url           = {https://arxiv.org/abs/1711.01053}
}

@article{HuangKuengPreskill2020,
  author  = {Huang, H.-Y. and Kueng, R. and Preskill, J.},
  title   = {Predicting many properties of a quantum system from very few measurements},
  journal = {Nat. Phys.},
  volume  = {16},
  pages   = {1050},
  year    = {2020},
  doi     = {10.1038/s41567-020-0932-7}
}

@article{Elben2023Toolbox,
  author  = {Elben, A. and Flammia, S. T. and Huang, H.-Y. and Kueng, R. and Preskill, J. and Vermersch, B. and Zoller, P.},
  title   = {The randomized measurement toolbox},
  journal = {Nat. Rev. Phys.},
  volume  = {5},
  pages   = {9},
  year    = {2023},
  doi     = {10.1038/s42254-022-00535-2}
}

@article{Smacchia2011ClusterIsing,
  author  = {Smacchia, Pietro and Amico, Luigi and Facchi, Paolo and Fazio, Rosario and Florio, Giuseppe and Pascazio, Saverio and Vedral, Vlatko},
  title   = {Statistical mechanics of the cluster {I}sing model},
  journal = {Phys. Rev. A},
  volume  = {84},
  pages   = {022304},
  year    = {2011},
  doi     = {10.1103/PhysRevA.84.022304}
}

@article{chaotic_ising,
  title = {Ballistic Spreading of Entanglement in a Diffusive Nonintegrable System},
  author = {Kim, Hyungwon and Huse, David A.},
  journal = {Phys. Rev. Lett.},
  volume = {111},
  issue = {12},
  pages = {127205},
  numpages = {5},
  year = {2013},
  month = {Sep},
  publisher = {American Physical Society},
  doi = {10.1103/PhysRevLett.111.127205},
  url = {https://link.aps.org/doi/10.1103/PhysRevLett.111.127205}
}

@article{PhysRevLett.106.050405,
  title = {Strong and Weak Thermalization of Infinite Nonintegrable Quantum Systems},
  author = {Ba{\~n}uls, M. C. and Cirac, J. I. and Hastings, M. B.},
  journal = {Phys. Rev. Lett.},
  volume = {106},
  issue = {5},
  pages = {050405},
  numpages = {4},
  year = {2011},
  month = {Feb},
  publisher = {American Physical Society},
  doi = {10.1103/PhysRevLett.106.050405},
  url = {https://link.aps.org/doi/10.1103/PhysRevLett.106.050405}
}

@article{Popescu2006,
  author  = {Popescu, S. and Short, A. J. and Winter, A.},
  title   = {Entanglement and the foundations of statistical mechanics},
  journal = {Nat. Phys.},
  volume  = {2},
  pages   = {754},
  year    = {2006},
  doi     = {10.1038/nphys444}
}

@article{Goldstein2006,
  author  = {Goldstein, S. and Lebowitz, J. L. and Tumulka, R. and Zangh{\`i}, N.},
  title   = {Canonical Typicality},
  journal = {Phys. Rev. Lett.},
  volume  = {96},
  pages   = {050403},
  year    = {2006},
  doi     = {10.1103/PhysRevLett.96.050403}
}

@book{Williams1991,
  author    = {Williams, D.},
  title     = {Probability with Martingales},
  publisher = {Cambridge University Press},
  address   = {Cambridge},
  year      = {1991},
  doi       = {10.1017/CBO9780511813658}
}

@article{GrossNezamiWalter2021,
  author        = {Gross, David and Nezami, Sepehr and Walter, Michael},
  title         = {{Schur--Weyl} Duality for the {Clifford} Group with Applications: Property Testing, a Robust {Hudson} Theorem, and de {Finetti} Representations},
  journal       = {Communications in Mathematical Physics},
  volume        = {385},
  pages         = {1325--1393},
  year          = {2021},
  doi           = {10.1007/s00220-021-04118-7},
  eprint        = {1712.08628},
  archivePrefix = {arXiv},
  primaryClass  = {quant-ph}
}

@article{Zhu2017Clifford3Design,
  author        = {Zhu, Huangjun},
  title         = {Multiqubit {Clifford} groups are unitary 3-designs},
  journal       = {Physical Review A},
  volume        = {96},
  pages         = {062336},
  year          = {2017},
  doi           = {10.1103/PhysRevA.96.062336},
  eprint        = {1510.02619},
  archivePrefix = {arXiv},
  primaryClass  = {quant-ph}
}

@article{ShendeBullockMarkov2006Synthesis,
  author        = {Shende, Vivek V. and Bullock, Stephen S. and Markov, Igor L.},
  title         = {Synthesis of quantum logic circuits},
  journal       = {IEEE Trans. Comput.-Aided Des. Integr. Circuits Syst.},
  volume        = {25},
  number        = {6},
  pages         = {1000--1010},
  year          = {2006},
  doi           = {10.1109/TCAD.2005.855930},
  eprint        = {quant-ph/0406176},
  archivePrefix = {arXiv}
}

@article{Webb2016Clifford3Design,
  author  = {Webb, Zak},
  title   = {The {Clifford} group forms a unitary 3-design},
  journal = {Quantum Inf. Comput.},
  volume  = {16},
  number  = {15--16},
  pages   = {1379--1400},
  year    = {2016},
  doi     = {10.26421/QIC16.15-16-8}
}

@misc{KuengGross2015Stabilizer3Design,
  author        = {Kueng, Richard and Gross, David},
  title         = {Qubit stabilizer states are complex projective 3-designs},
  year          = {2015},
  eprint        = {1510.02767},
  archivePrefix = {arXiv},
  primaryClass  = {quant-ph},
  doi           = {10.48550/arXiv.1510.02767},
  url           = {https://arxiv.org/abs/1510.02767}
}

@article{AaronsonGottesman2004StabilizerCircuits,
  author  = {Aaronson, Scott and Gottesman, Daniel},
  title   = {Improved simulation of stabilizer circuits},
  journal = {Phys. Rev. A},
  volume  = {70},
  pages   = {052328},
  year    = {2004},
  doi     = {10.1103/PhysRevA.70.052328}
}

@article{CollinsSniady2006,
  author        = {Collins, Beno{\^i}t and {\'S}niady, Piotr},
  title         = {Integration with Respect to the {Haar} Measure on Unitary, Orthogonal and Symplectic Group},
  journal       = {Commun. Math. Phys.},
  volume        = {264},
  pages         = {773--795},
  year          = {2006},
  doi           = {10.1007/s00220-006-1554-3},
  eprint        = {math-ph/0402073},
  archivePrefix = {arXiv}
}

@misc{FarhiGoldstoneGutmann2014,
  author        = {Farhi, Edward and Goldstone, Jeffrey and Gutmann, Sam},
  title         = {A Quantum Approximate Optimization Algorithm},
  year          = {2014},
  eprint        = {1411.4028},
  archivePrefix = {arXiv},
  primaryClass  = {quant-ph},
  doi           = {10.48550/arXiv.1411.4028},
  url           = {https://arxiv.org/abs/1411.4028}
}

@article{Biamonte2017QML,
  author  = {Biamonte, Jacob and Wittek, Peter and Pancotti, Nicola and Rebentrost, Patrick and Wiebe, Nathan and Lloyd, Seth},
  title   = {Quantum machine learning},
  journal = {Nature},
  volume  = {549},
  pages   = {195--202},
  year    = {2017},
  doi     = {10.1038/nature23474}
}

@misc{Huang2025GenerativeQuantumAdvantage,
  author        = {Huang, Hsin-Yuan and Broughton, Michael and Eassa, Norhan and Neven, Hartmut and Babbush, Ryan and McClean, Jarrod R.},
  title         = {Generative quantum advantage for classical and quantum problems},
  year          = {2025},
  eprint        = {2509.09033},
  archivePrefix = {arXiv},
  primaryClass  = {quant-ph},
  doi           = {10.48550/arXiv.2509.09033},
  url           = {https://arxiv.org/abs/2509.09033}
}

@article{LloydWeedbrook2018,
  author  = {Lloyd, Seth and Weedbrook, Christian},
  title   = {Quantum Generative Adversarial Learning},
  journal = {Phys. Rev. Lett.},
  volume  = {121},
  pages   = {040502},
  year    = {2018},
  doi     = {10.1103/PhysRevLett.121.040502}
}

@article{Cong2019QCNN,
  author  = {Cong, Iris and Choi, Soonwon and Lukin, Mikhail D.},
  title   = {Quantum convolutional neural networks},
  journal = {Nature Physics},
  volume  = {15},
  pages   = {1273--1278},
  year    = {2019},
  doi     = {10.1038/s41567-019-0648-8}
}

@article{Gao2022GenerativeCorrelations,
  author  = {Gao, Xun and Anschuetz, Eric R. and Wang, Sheng-Tao and Cirac, J. Ignacio and Lukin, Mikhail D.},
  title   = {Enhancing Generative Models via Quantum Correlations},
  journal = {Phys. Rev. X},
  volume  = {12},
  pages   = {021037},
  year    = {2022},
  doi     = {10.1103/PhysRevX.12.021037}
}

@article{Zhang2024QuantumDiffusion,
  author  = {Zhang, Bingzhi and Xu, Peng and Chen, Xiaohui and Zhuang, Quntao},
  title   = {Generative Quantum Machine Learning via Denoising Diffusion Probabilistic Models},
  journal = {Phys. Rev. Lett.},
  volume  = {132},
  pages   = {100602},
  year    = {2024},
  doi     = {10.1103/PhysRevLett.132.100602}
}

@misc{Liu2025MeasurementBasedQuantumDiffusion,
  author        = {Liu, Xinyu and Zhuang, Jingze and Hou, Wanda and You, Yi-Zhuang},
  title         = {Measurement-Based Quantum Diffusion Models},
  year          = {2025},
  eprint        = {2508.08799},
  archivePrefix = {arXiv},
  primaryClass  = {quant-ph},
  doi           = {10.48550/arXiv.2508.08799},
  url           = {https://arxiv.org/abs/2508.08799}
}

@article{Clemente2020ThermalAverages,
  author        = {Clemente, Giuseppe and Cardinali, Marco and Bonati, Claudio and Calore, Enrico and Cosmai, Leonardo and D'Elia, Massimo and Gabbana, Alessandro and Rossini, Davide and Schifano, Fabio Sebastiano and Tripiccione, Raffaele and Vadacchino, Davide},
  collaboration = {QuBiPF},
  title         = {Quantum computation of thermal averages in the presence of a sign problem},
  journal       = {Phys. Rev. D},
  volume        = {101},
  pages         = {074510},
  year          = {2020},
  doi           = {10.1103/PhysRevD.101.074510},
  eprint        = {2001.05328},
  archivePrefix = {arXiv},
  primaryClass  = {hep-lat}
}

@article{Yamamoto2022QuantumSampling,
  author        = {Yamamoto, Arata},
  title         = {Quantum sampling for the {Euclidean} path integral of lattice gauge theory},
  journal       = {Phys. Rev. D},
  volume        = {105},
  pages         = {094501},
  year          = {2022},
  doi           = {10.1103/PhysRevD.105.094501},
  eprint        = {2201.12556},
  archivePrefix = {arXiv},
  primaryClass  = {quant-ph}
}

@article{ChildsWiebe2013Commutators,
  author  = {Childs, Andrew M. and Wiebe, Nathan},
  title   = {Product formulas for exponentials of commutators},
  journal = {Journal of Mathematical Physics},
  volume  = {54},
  number  = {6},
  pages   = {062202},
  year    = {2013},
  doi     = {10.1063/1.4811386}
}

@unpublished{WangChowdhuryShadowGeometry2026,
  author = {Wang, Xuepeng and Chowdhury, Debanjan},
  title  = {Controlling Randomness in Quantum Shadow Chains},
  note   = {Unpublished companion manuscript},
  year   = {2026}
}
\clearpage
\setcounter{section}{0}
\setcounter{equation}{0}
\setcounter{figure}{0}
\setcounter{table}{0}
\renewcommand{\theequation}{S\arabic{equation}}
\renewcommand{\thefigure}{S\arabic{figure}}
\renewcommand{\thetable}{S\arabic{table}}
\renewcommand{\figurename}{Supplemental Figure}
\renewcommand{\theHsection}{supp.S.\arabic{section}}
\renewcommand{\theHequation}{supp.S.\arabic{equation}}
\renewcommand{\theHfigure}{supp.S.\arabic{figure}}
\renewcommand{\theHtable}{supp.S.\arabic{table}}
\providecommand{\theHtheorem}{}
\renewcommand{\theHtheorem}{supp.S.\arabic{section}.\arabic{theorem}}
\onecolumngrid
\begin{center}
\Large{\textbf{Supplementary Material for} ``Quantum Cooling by Classical Shadows''}\\[4pt]
\large{X. Wang and D. Chowdhury}
\end{center}
In this Supplementary Material, we provide additional technical details regarding the shadow chain protocol, which is briefly described in Algorithm~\ref{alg:shadow_chain}.
\begin{algorithm}[htb!]
\caption{Shadow Chain}
\label{alg:shadow_chain}
\small
\SetInd{0.5em}{1em}
\SetNlSkip{0.6em}
\DontPrintSemicolon
\KwIn{Hamiltonian $h$ on a $d$-dimensional register; preparation circuit for $\ket{\psi_0}$; pulse sizes $\alpha,\theta>0$; number of steps $N_T$.}
\KwOut{Trajectory state $\ket{\psi_{N_T}}$ and classical shadow history $\Omega_{N_T}$.}
Set $\eta\gets\alpha\theta/(d+1)$ and $\Omega_0\gets\emptyset$\;
\For{$k=0,\ldots,N_T-1$}{
    Prepare $\ket{\psi_0}$ and replay $V_0,\ldots,V_{k-1}$ to obtain $\ket{\psi_k}$ (empty replay for $k=0$)\;
    Draw $U_k$ independently from the Haar measure\;
    Apply $U_k^\dagger$ and measure in the computational basis, collapsing the state to $\ket{b_k}$\;
    Store $\omega_k\gets\{U_k,b_k\}$ and set $\Omega_{k+1}\gets(\Omega_k,\omega_k)$\;
    Set $P_k\gets U_k\ket{b_k}\bra{b_k}U_k^\dagger$\;
    Define $V_k\gets e^{i\alpha h}e^{i\theta P_k}e^{-i\alpha h}e^{-i\theta P_k}$\;
}
Prepare $\ket{\psi_0}$ and replay $V_0,\ldots,V_{N_T-1}$ to obtain $\ket{\psi_{N_T}}$\;
\Return{$\ket{\psi_{N_T}}$ and $\Omega_{N_T}$}\;
\end{algorithm}

The remainder of this material is organized as follows. Section~\ref{sec:qite} introduces normalized imaginary-time evolution (ITE) and the thermofield-double reference. Section~\ref{sec:sde} decomposes the physical gate update into the ITE drift, martingale difference, and BCH and Euler errors. Section~\ref{sec:supp_moving_frame} derives the exact propagation and accumulation of these sources, together with bounds on the thermal propagator. Section~\ref{sec:supp_bch_error} establishes the local source moment bounds and controls the accumulated normalization changes.
These estimates give the observable-bias and trajectory-variance bounds in Secs.~\ref{sec:supp_gate_bias} and~\ref{sec:supp_gate_variance}, including the thermal-observable and state-distance corollaries. Section~\ref{sec:supp_mom} gives the median-of-means estimation guarantee for independent trajectories. Finally, Sec.~\ref{sec:supp_additional_benchmarks} presents additional benchmarks for the Ising and Heisenberg chains.

The physical system has dimension $N=2^L$, while $d$ denotes the dimension of the measured register. For thermofield-double (TFD) preparation, the register contains the system and an ancilla, so $d=N^2$ and $h=H_{\rm sys}\otimes\mathbb1_{\rm anc}$. We use $\|\cdot\|_{\rm op}$ and $\|\cdot\|_1$ for the operator and trace norms. The conditional expectation $\mathbb E_i$ is taken over the next measurement, with the preceding shadow history $\Omega_i$ held fixed.

\section{Imaginary-time reference}\label{sec:qite}
The continuous reference flow generated by the normalized ITE map is defined as
\be
\label{eq:dm_ite_suppl_pure}
\ket{\phi_\tau}=\frac{e^{-\tau h}\ket{\phi_0}}{\|e^{-\tau h}\ket{\phi_0}\|}, \qquad \dot\rho_\phi=\mathcal F(\rho_\phi),\qquad \mathcal F(X)=-\{h,X\}+2\tn{Tr}(hX)X.
\ee
For a pure state, this reduces to the double-bracket flow $\mathcal F(\rho)=F(\rho)\equiv[[\rho,h],\rho]$~\cite{Gluza2026DBQITE}; the generator $\mathcal F$ allows us to propagate a general density-matrix deviation.

We initialize both the reference and the shadow chain in $\ket{\psi_0}=\ket{\phi_0}=N^{-1/2}\sum_a\ket a_{\rm sys}\ket a_{\rm anc}$. The reference evolved by Eq.~\eqref{eq:dm_ite_suppl_pure} with $h=H\otimes \mathbb{1}$ is then a TFD state at inverse temperature $\beta=2\tau$. Consequently, its expectation of $O=O_{\rm sys}\otimes\mathbb1_{\rm anc}$ is the thermal expectation of $O_{\rm sys}$. We denote the system partition function by $\mathcal{Z}(\beta)=\tn{Tr}_{\rm sys}(e^{-\beta H_{\rm sys}})$.

\section{Physical gate and source decomposition}\label{sec:sde}\label{sec:doob_decomp}
Our goal is to analyze the density-matrix difference between the physical gate update and the continuous normalized ITE map in a single step (as detailed below).
At step $i$, the shadow measurement produces a rank-one projector $P_i=U_i\ket{b_i}\bra{b_i}U_i^\dagger$, where $\ket{b_i}$ is a computational basis measurement outcome and $U_i$ is a Haar-random unitary. The physical gate update is a four-pulse sequence defined as
\be
\label{eq:v_def_suppl}
V_i=e^{i\alpha h}e^{i\theta P_i}e^{-i\alpha h}e^{-i\theta P_i}, \qquad \rho_{\psi_{i+1}}=V_i\rho_{\psi_i}V_i^\dagger.
\ee
The pulse parameters satisfy $\alpha\theta>0$, with $\alpha$ in units of inverse energy and $\theta$ dimensionless. 
Since the conditional mean of the projector is $\mathbb{E}[P_i|\rho_{\psi_i}]=(\rho_{\psi_i}+\mathbb1)/(d+1)$, the leading physical update can be expressed as a discrete stochastic differential equation (SDE) of density matrices, with a deterministic drift term proportional to $F(\rho_{\psi_i})$ and a centered fluctuation $\Xi_{i+1}^{(\psi)}$ defined as,
\begin{subequations}
\beq
\label{eq:sde_dm_0}
\rho_{\psi_{i+1}}
&=&
\rho_{\psi_i}
+\eta F(\rho_{\psi_i})
+\Xi_{i+1}^{(\psi)}
+O(\alpha^2\theta,\alpha\theta^2),
\\
\label{eq:supp_matrix_noise_gen}
\Xi_{i+1}^{(\psi)} &=&\alpha\theta\left[\left[P_i-\frac{\rho_{\psi_i}+\mathbb1}{d+1},h\right],\rho_{\psi_i}\right], \qquad \mathbb E_i\Xi_{i+1}^{(\psi)}=0.
\eeq
\end{subequations}
In Eq.~\eqref{eq:sde_dm_0}, the step size is $\eta=\alpha\theta/(d+1)$ and the higher-order remainder contains BCH and Euler contributions as we will address below.
Since $\mathbb E_i\Xi_{i+1}^{(\psi)}=0$, these increments are martingale differences; their accumulated sum is a matrix-valued martingale with respect to the shadow history $\Omega_i = \{\omega_j\}_{j\le i}$ as a filtration in probability theory with $\omega_j\equiv \{\ket{b_j}, U_j \}$. The martingale differences defined in Eq.~\eqref{eq:supp_matrix_noise_gen} are traceless and satisfy an orthogonality condition at distinct steps.

To compare the physical gate update with the exact ITE map, let us introduce the generator of the reference continuous flow $\mathcal{T}_{f,i}$ as
\be
\label{eq:def_mathcal_T}
\mathcal T_{f,i}[\rho]=\frac{e^{-(\tau_f-\tau_i)h}\rho e^{-(\tau_f-\tau_i)h}}{\tn{Tr}(e^{-2(\tau_f-\tau_i)h}\rho)},
\ee
where $\tau_k=k\eta$ is a discrete grid of imaginary time with step size $\eta=\alpha\theta/(d+1)$. Since the generator in Eq.~\eqref{eq:def_mathcal_T} satisfies the composition rule $\mathcal T_{f,i}\circ\mathcal T_{i,j}=\mathcal T_{f,j}$, the reference density matrix $\rho_{\phi_i}$ evolved by this generator on the discrete grid is \textit{sampled} along the continuous reference flow rather than evolved via Trotterization.

The exact difference between the physical gate update and the normalized ITE map separates into BCH, Euler, and martingale-difference contributions as
\begin{subequations}
\label{eq:supp_exact_source_split}
\begin{align}
\label{eq:main_total_source}
J_{i+1}&=V_i\rho_{\psi_i}V_i^\dagger-\mathcal T_{i+1,i}[\rho_{\psi_i}]=J_{i+1}^{\rm BCH}+J_{i+1}^{\rm E}+\Xi_{i+1}^{(\psi)},\\
\label{eq:main_bch_source}
J_{i+1}^{\rm BCH}&=V_i\rho_{\psi_i}V_i^\dagger-\rho_{\psi_i}-\alpha\theta[[P_i,h],\rho_{\psi_i}],\\
\label{eq:main_euler_source}
J_{i+1}^{\rm E}&=\rho_{\psi_i}+\eta F(\rho_{\psi_i})-\mathcal T_{i+1,i}[\rho_{\psi_i}].
\end{align}
\end{subequations}
The BCH contribution can be separated further into its conditional mean $J_{i+1}^{\overline{\rm BCH}}=\mathbb E_iJ_{i+1}^{\rm BCH}$ and centered fluctuation $J_{i+1}^{\widetilde{\rm BCH}}=J_{i+1}^{\rm BCH}-J_{i+1}^{\overline{\rm BCH}}$. 
These decompositions allow us to separate the source terms into conditionally deterministic drift terms, $J_{i+1}^{\overline{\rm BCH}}$ and $J_{i+1}^{\rm E}$, and centered fluctuations, $J_{i+1}^{\widetilde{\rm BCH}}$ and $\Xi_{i+1}^{(\psi)}$, respectively. We treat the shadow evolution in Eq.~\eqref{eq:v_def_suppl} as a driven problem with the total source $J_{i+1}$ in Eq.~\eqref{eq:main_total_source}, so that the physical-gate update is equivalent to the exact recursion
\be
\label{eq:main_driven_update}
\rho_{\psi_{i+1}}=\mathcal T_{i+1,i}[\rho_{\psi_i}]+J_{i+1}.
\ee
The source $J_{i+1}$ is injected at the end of the interval $[\tau_i,\tau_{i+1}]$ after the normalized ITE propagation, which we refer to as a post-kick injection. Since the centered fluctuations have zero conditional mean, they correspond to jumps between different flows after each step, as schematically illustrated in Figure~\ref{Fig:flows}(a); the deterministic drift terms bias the flow away from the normalized ITE map on top of these jumps, as illustrated in Figure~\ref{Fig:flows}(b).

\ifdefined\supptextonly\refstepcounter{figure}\label{Fig:flows}\else
\begin{figure*}[htb!]
\centering
\includegraphics[width=15cm]{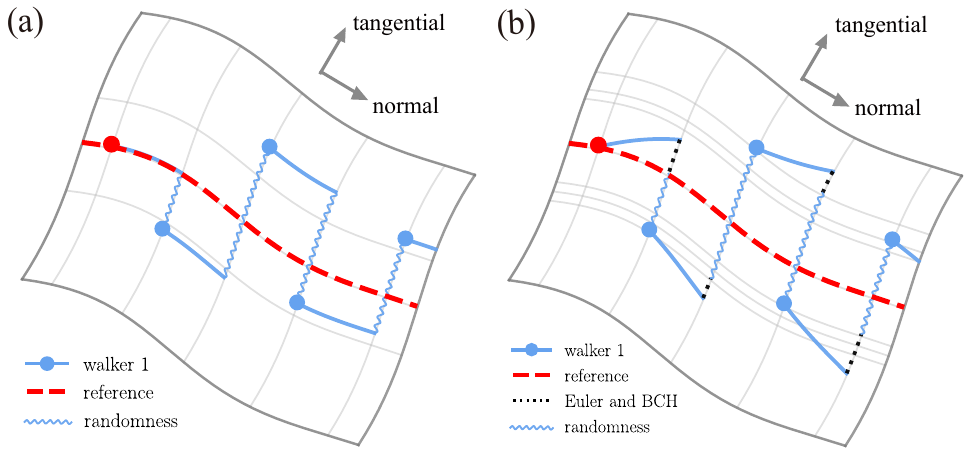}
\caption{\label{Fig:flows} {\bf Propagation of sample differences.} {\bf a} Idealized stochastic flow, which follows the normalized ITE map between steps and receives a randomness injection due to shadow measurement (blue wavy lines). {\bf b} Physical gate evolution. Relative to the normalized ITE map (blue solid lines), the gate generates the centered shadow randomness (blue wavy lines) and the BCH and Euler bias (black dotted lines). The exact recursion propagates and accumulates these sources to late time $\tau_k$, giving rise to the total sample difference $Z_k\equiv \rho_{\psi_k}-\rho_{\phi_k}$.}
\end{figure*}
\fi

\section{Exact propagation of accumulated errors}\label{sec:supp_moving_frame}

\subsection{Propagating a generic density-matrix perturbation}

Having demonstrated the sources of systematic deviation generated per step in Eq.~\eqref{eq:supp_exact_source_split}, we now propagate them along the physical trajectory. Let us write 
\be
\label{eq:suppl_def_z}
Z_i=\rho_{\psi_i}-\rho_{\phi_i},
\ee
and define the propagator normalized with respect to the reference state,
\be
\label{eq:def_map_gf}
G_{f,i}^{\phi}=e^{-(\tau_f-\tau_i)h}\sqrt{\frac{\mathcal{Z}(2\tau_i)}{\mathcal{Z}(2\tau_f)}},
\ee
where $\mathcal{Z}(\tau)\equiv \tn{Tr}_{\tn{sys}}(e^{-\tau H_{\tn{sys}}})$ is the physical partition function.
The flow generator $\mathcal{T}_{f,i}$ can be expressed in terms of the Green's function, which also satisfies the composition rule:
\be
\label{eq:main_propagator_composition}
\mathcal{T}_{f,i}[\rho_{\phi_i}]
=
G_{f,i}^{\phi}\rho_{\phi_i}G_{f,i}^{\phi}=\rho_{\phi_f},\qquad G_{f,i}^{\phi}G_{i,j}^{\phi}=G_{f,j}^{\phi}\quad(0\leq j\leq i\leq f).
\ee

Suppose that, at step $i$, the sampled state is $\rho_{\psi_i}$. We define the propagation of this deviation to a later step $f>i$ as $D_{f,i}[Z]$, which is given by
\be
\label{eq:exact_sol_z_dm}
D_{f,i}[Z]\equiv\mathcal T_{f,i}[\rho_{\phi_i}+Z]-\rho_{\phi_f} =\frac{G_{f,i}^{\phi}ZG_{f,i}^{\phi}-\rho_{\phi_f}\tn{Tr}(G_{f,i}^{\phi}ZG_{f,i}^{\phi})} {1+\tn{Tr}(G_{f,i}^{\phi}ZG_{f,i}^{\phi})}.
\ee
Since $\rho_{\psi_i}$ is also a density matrix related to $\rho_{\phi_i}$ via Eq.~\eqref{eq:suppl_def_z}, the denominator equals $\tn{Tr}[G_{f,i}^{\phi}(\rho_{\phi_i}+Z)G_{f,i}^{\phi}]>0$. Equation~\eqref{eq:exact_sol_z_dm} is exact and includes the non-linearity in the normalization change at every order in $Z$.

\subsection{Accumulated sample differences during propagation}
With the source convention in Eq.~\eqref{eq:supp_exact_source_split}, the deviation satisfies $Z_i=D_{i,i-1}[Z_{i-1}]+J_i$, where the new source $J_i$ is added after propagating $Z_{i-1}$ to step $i$. Starting from $X_0=Z_0=J_0=0$, we obtain the exact solution
\begin{subequations}
\label{eq:sol_z_recursive}
\begin{align}
\label{eq:main_recursive_history}
X_k&=\sum_{1\leq i<k}(1+r_i)G_{k,i}^{\phi}J_iG_{k,i}^{\phi},\qquad r_k=\tn{Tr}X_k,\\
\label{eq:main_recursive_deviation}
Z_k&=\frac{X_k-r_k\rho_{\phi_k}}{1+r_k}+J_k.
\end{align}
\end{subequations}
The weights $1+r_i$ account for the normalization changes produced by earlier sources. The terminal source $J_k$ appears separately because it has no remaining propagation interval.

We prove Eq.~\eqref{eq:sol_z_recursive} by induction. The initial condition $X_0=Z_0=J_0=r_0=0$ establishes the base case. Suppose the result holds at step $i$. Using the relation between the sampled state $\rho_{\psi_i}$ and the reference state $\rho_{\phi_i}$ in Eq.~\eqref{eq:suppl_def_z} and multiplying Eq.~\eqref{eq:main_recursive_deviation} by $1+r_i$ gives
\be
\label{eq:main_homogeneous_state}
(1+r_i)\rho_{\psi_i}=\rho_{\phi_i}+X_i+(1+r_i)J_i.
\ee
Comparing $X_{i+1}$ with $X_{i}$ using Eq.~\eqref{eq:main_recursive_history} gives the following relation
\begin{subequations}
\label{eq:main_homogeneous_recursion}
\be
\label{eq:main_homogeneous_matrix}
X_{i+1}=G_{i+1,i}^{\phi}X_iG_{i+1,i}^{\phi}+(1+r_i)G_{i+1,i}^{\phi}J_iG_{i+1,i}^{\phi}.
\ee
A similar relation between $r_{i+1}$ and $r_{i}$ can be obtained by multiplying Eq.~\eqref{eq:main_homogeneous_state} on both sides by $G_{i+1,i}^{\phi}$ and using Eq.~\eqref{eq:main_homogeneous_matrix} as
\be
\label{eq:main_propagated_state}
(1+r_i)G_{i+1,i}^{\phi}\rho_{\psi_i}G_{i+1,i}^{\phi}=\rho_{\phi_{i+1}}+X_{i+1},
\ee
and by taking the trace of Eq.~\eqref{eq:main_propagated_state}, we obtain
\be
\label{eq:main_homogeneous_normalization}
(1+r_i)\tn{Tr}(G_{i+1,i}^{\phi}\rho_{\psi_i}G_{i+1,i}^{\phi})=1+r_{i+1}.
\ee
\end{subequations}
Dividing Eq.~\eqref{eq:main_propagated_state} by Eq.~\eqref{eq:main_homogeneous_normalization} gives the normalized ITE state, $\mathcal T_{i+1,i}[\rho_{\psi_i}]=(\rho_{\phi_{i+1}}+X_{i+1})/(1+r_{i+1})$. By the definition of $J_{i+1}$, the next physical state is this normalized state plus $J_{i+1}$. Subtracting $\rho_{\phi_{i+1}}$ then gives $Z_{i+1}=(X_{i+1}-r_{i+1}\rho_{\phi_{i+1}})/(1+r_{i+1})+J_{i+1}$, which is Eq.~\eqref{eq:main_recursive_deviation} at the next step. This completes the induction.
Since $r_0=0$ and the physical states are density matrices, Eq.~\eqref{eq:main_homogeneous_normalization} also ensures that all $1+r_i$ are positive.

\subsection{The constituent relation}
The recursive solution allows us to relate the information at time $i$ to the fresh martingale contributions from all previous times $j\leq i$, and we may further propagate it to a later time $k$ by applying the Green's function $G_{k,i}^{\phi}$.
After propagating Eq.~\eqref{eq:main_homogeneous_state} to $k$ and taking the trace, its right-hand side becomes $1+\tn{Tr}(G_{k,i}^{\phi}X_iG_{k,i}^{\phi})+(1+r_i)\tn{Tr}(G_{k,i}^{\phi}J_iG_{k,i}^{\phi})$. Substituting the history sum for $X_i$ yields the constituent relation
\begin{subequations}
\beq
s_{k,i}&\equiv&\tn{Tr}(G_{k,i}^{\phi}\rho_{\psi_i}G_{k,i}^{\phi})
\\
b_{k,j}&\equiv&\tn{Tr}(G_{k,j}^{\phi}J_jG_{k,j}^{\phi})
\\
\label{eq:relate_rsb_ideal}
(1+r_i)s_{k,i}-1&=&\sum_{j=1}^{i}(1+r_j)b_{k,j},\qquad 0\leq i\leq k.
\eeq
\end{subequations}
This relation is crucial for deriving the bounds for the bias and variance below.

\subsection{Bounds from the TFD state}\label{sec:supp_tfd_propagator_bounds}
For TFD initialization, the amplification of a source is bounded uniformly in the propagation time. The eigenvalues of $G_{f,i}^{\phi}$ are $e^{-(\tau_f-\tau_i)E_a}\sqrt{\mathcal{Z}(2\tau_i)/\mathcal{Z}(2\tau_f)}$, each repeated $N$ times by the ancilla. The factors $e^{-2\tau_iE_a}$ and $e^{-2(\tau_f-\tau_i)E_a}$ have the same ordering with energy, so Chebyshev's sum inequality gives,
\be
\label{eq:main_partition_ordering}
\mathcal{Z}(2\tau_i)\mathcal{Z}(2\tau_f-2\tau_i)\leq N\mathcal{Z}(2\tau_f),\qquad \tn{Tr}[(G_{f,i}^{\phi})^2] =N\frac{\mathcal{Z}(2\tau_i)\mathcal{Z}(2\tau_f-2\tau_i)}{\mathcal{Z}(2\tau_f)}\leq d.
\ee
A common shift of the energies leaves $G_{f,i}^{\phi}$ unchanged. Choosing the minimum energy to be zero gives $\mathcal{Z}(2\tau_i)\leq N$, $\mathcal{Z}(2\tau_f)\geq1$, and $e^{-2(\tau_f-\tau_i)E_a}\leq1$, so every squared eigenvalue is at most $N$. Finally, $\tn{Tr}(G^4)\leq\|G\|_{\rm op}^2\tn{Tr}(G^2)$. Together these estimates give
\be
\label{eq:supp_green_bounds}
\tn{Tr}[(G_{f,i}^{\phi})^2]\leq d, \qquad \|G_{f,i}^{\phi}\|_{\rm op}^2\leq N, \qquad \tn{Tr}[(G_{f,i}^{\phi})^4]\leq N^3.
\ee
These bounds hold for every $0\leq i\leq f$ and depend only on the register dimension, rather than growing exponentially with the elapsed imaginary time.

\section{Intrinsic controlled parameter}\label{sec:supp_bch_error}
The constituent relation, Eq.~\eqref{eq:relate_rsb_ideal}, relates the normalization change $s_{k,i}$ at late time to the weighted source contributions $(1+r_j)b_{f,j}$ for $j\leq i$.
Importantly, $b_{f,j}$ depends on the source terms $J_j^{(\#)}$ defined in Eq.~\eqref{eq:supp_exact_source_split}, which can be upper bounded in terms of $s_{f,j-1}$ upon conditional averaging (since $\mathbb E_{j-1}[b_{f,j}^{(\#)}]$ can involve $\rho_{\psi_{j-1}}$; a similar form holds for higher powers of $b_{f,j}^{(\#)}$). This enables us to obtain a closed set of inequalities with a controlled parameter and is the key to proving the bias and variance bounds in the main text.

Throughout this section, $1\leq j\leq f\leq k$, and the source $J_j$ is evaluated on the actual input state $\rho_{\psi_{j-1}}$, held fixed under the conditional expectation $\mathbb E_{j-1}$. The observable bounds use a deterministic Hermitian matrix $B$.

\subsection{Local source moments}\label{subsec:local_source_bias}
Let us separate the contributions to $b_{f,j}$ from the four sources $\#\in\{\overline{\rm BCH},{\rm E},\widetilde{\rm BCH},\Xi\}$, with $J_j^{(\Xi)}=\Xi_j^{(\psi)}$, and write
\be
\label{eq:main_source_components}
b_{f,j}^{(\#)}=\tn{Tr}(G_{f,j}^{\phi}J_j^{(\#)}G_{f,j}^{\phi}),\qquad b_{f,j}=\sum_\#b_{f,j}^{(\#)}.
\ee
For the pure state $\rho_{\psi_{j-1}}$, with nonnegative coefficients $A_\alpha,C_\alpha,\widetilde A_\alpha,\widetilde C_\alpha$, $\lambda_{\rm E}$ and $\lambda_\Xi$ obtained by evaluating the Born-weighted Haar contractions, the contributions relevant to $(1+r_j)b_{f,j}$ in the constituent relation obey the following bounds
\begin{subequations}
\label{eq:main_source_bounds}
\begin{align}
\label{eq:main_drift_source_bound}
\left|(1+r_j)b_{f,j}^{(\#)}\right|&\leq
\begin{cases}
A_\alpha(1+r_{j-1})s_{f,j-1}+C_\alpha(1+r_j),&\#=\overline{\rm BCH},\\
\lambda_{\rm E}(1+r_j),&\#={\rm E},
\end{cases}\\
\label{eq:main_centered_source_bound}
\mathbb E_{j-1}\!\left[(1+r_j)^2\big(b_{f,j}^{(\#)}\big)^2\right]&\leq
\begin{cases}
\widetilde A_\alpha(1+r_{j-1})^2s_{f,j-1}+\widetilde C_\alpha(1+r_j)^2,&\#=\widetilde{\rm BCH},\\
\lambda_\Xi(1+r_{j-1})^2s_{f,j-1},&\#=\Xi.
\end{cases}
\end{align}
\end{subequations}
Equation~\eqref{eq:main_drift_source_bound} controls the two drift sources, which are constant under $\mathbb E_{j-1}$; equation~\eqref{eq:main_centered_source_bound} controls the conditional second moments of the centered sources, for which $\mathbb E_{j-1}b_{f,j}^{(\#)}=0$. Replacing each $b_{f,j}^{(\#)}$ by $\tn{Tr}(BG_{f,j}^{\phi}J_j^{(\#)}G_{f,j}^{\phi})$ gives the same inequalities with the right-hand sides multiplied by $\|B\|_{\rm op}$ and $\|B\|_{\rm op}^2$, respectively.

Let us briefly sketch the procedure to obtain the  coefficients in Eq.~\eqref{eq:main_source_bounds}. 
Defining $\mathcal W=V_{j-1}-\mathbb1$ and $\mathcal E=\mathcal W-\alpha\theta[P_{j-1},h]$, the BCH source has the exact decomposition 
$
J_j^{\rm BCH}=\mathcal E\rho_{\psi_{j-1}}+\rho_{\psi_{j-1}}\mathcal E^\dagger +\mathcal W\rho_{\psi_{j-1}}\mathcal W^\dagger.
$
The identity $e^{i\theta P}=\mathbb1+(e^{i\theta}-1)P$ and the fact that $\rho_{\psi}$ is pure reduce all contributions to finite sums of projector contractions, which give rise to the following small parameters
\be
\label{eq:bch_aggregate_envelopes}
\mathcal B_{\mathcal W}=2|\alpha\theta|\|h\|_{\rm op}(1+|\theta|),\qquad \mathcal B_{\mathcal E}=|\alpha\theta|\|h\|_{\rm op}(3|\theta|+2|\alpha|\|h\|_{\rm op}).
\ee
These follow from $|e^{i\theta}-1|\leq|\theta|$, $\|e^{i\alpha h}-\mathbb1\|_{\rm op}\leq|\alpha|\|h\|_{\rm op}$, and the corresponding second-order remainder inequalities. Applying the Born-weighted Haar contractions gives
\begin{subequations}
\label{eq:main_gate_coefficients}
\begin{align}
\label{eq:main_gate_mean_coefficients}
A_\alpha&=\frac{4\mathcal B_{\mathcal E}}{d+1},\qquad C_\alpha=\frac{6d\mathcal B_{\mathcal W}^2}{(d+1)(d+2)},\\
\label{eq:main_gate_centered_coefficients}
\widetilde A_\alpha&=\frac{48N\mathcal B_{\mathcal E}^2}{(d+1)(d+2)},\qquad \widetilde C_\alpha=\frac{240d^2\mathcal B_{\mathcal W}^4}{\prod_{\ell=1}^{4}(d+\ell)}.
\end{align}
\end{subequations}
These are the BCH coefficients in Eq.~\eqref{eq:main_source_bounds}.

The Euler and martingale difference contributions have coefficients given by
\be
\label{eq:main_euler_shadow_rates}
\lambda_{\rm E}=12\eta^2N\|h\|_{\rm op}^2,\qquad \lambda_\Xi=\frac{24\alpha^2\theta^2N\|h\|_{\rm op}^2}{(d+1)(d+2)}.
\ee
The Euler coefficient follows from Taylor's integral remainder bound $\|J_j^{\rm E}\|_1\leq12\eta^2\|h\|_{\rm op}^2$. Propagation increases this bound by at most $N\|B\|_{\rm op}$, using the bounds for the TFD Green's function in Eq.~\eqref{eq:supp_green_bounds}; the martingale difference coefficient follows from the third Haar moment bound on $\Xi_j^{(\psi)}$.

\subsection{Closing the normalization estimate}

Given that the RHS of the constituent relation, Eq.~\eqref{eq:relate_rsb_ideal}, can be bounded by Eq.~\eqref{eq:main_source_bounds}, whose terms are proportional to polynomials of $s_{f,i}$, it is useful to define the controlling parameter as
\be
\label{eq:main_normalization_moment}
\mathcal S_k=\max_{0\leq i\leq f\leq k} \sqrt{\mathbb E\!\left[((1+r_i)s_{f,i}-1)^2\right]}.
\ee
Since $s_{i,i}=1$, Minkowski's inequality bounds both $\sqrt{\mathbb E[(1+r_i)^2]}$ and $\sqrt{\mathbb E[(1+r_i)^2s_{f,i}^2]}$ by $1+\mathcal S_k$ from above. Cauchy--Schwarz therefore also gives
\be
\label{eq:main_weight_moment_bound}
\mathbb E[(1+r_i)^2s_{f,i}] \leq\sqrt{\mathbb E[(1+r_i)^2]\,\mathbb E[(1+r_i)^2s_{f,i}^2]} \leq(1+\mathcal S_k)^2.
\ee
Taking expectations in Eq.~\eqref{eq:main_source_bounds} with $B$ inserted in each trace and applying the inequalities in Eq.~\eqref{eq:main_weight_moment_bound} gives the following bound on the RHS of the constituent relation
\be
\label{eq:main_weighted_source_moments}
\sqrt{\mathbb E\!\left[(1+r_j)^2\tn{Tr}(BG_{f,j}^{\phi}J_j^{(\#)}G_{f,j}^{\phi})^2\right]}
\leq(1+\mathcal S_k)\|B\|_{\rm op}
\begin{cases}
\lambda_\#,&\#\in\{\overline{\rm BCH},{\rm E}\},\\
\sqrt{\lambda_\#},&\#\in\{\widetilde{\rm BCH},\Xi\},
\end{cases}
\ee
where we have also used Minkowski's inequality for the drift sources.
The coefficients in Eq.~\eqref{eq:main_weighted_source_moments} are collected as
\beq
\label{eq:main_bch_rates}
\lambda_{\overline{\rm BCH}}=A_\alpha+C_\alpha,\qquad \lambda_{\widetilde{\rm BCH}}=\widetilde A_\alpha+\widetilde C_\alpha.
\eeq

We now bound the sum of Eq.~\eqref{eq:main_weighted_source_moments} over different initial steps $j$ for all the source terms $\#$. For each drift sum, Minkowski's inequality bounds its root second moment by the sum of the individual root second moments. For each centered source, $1+r_j$ is fixed before source $j$ is drawn, so $\mathbb E_{j-1}[(1+r_j)b_{f,j}^{(\#)}]=0$. Cross terms between distinct steps therefore vanish in the second moment of the sum. Consequently, for $0\leq i\leq f\leq k$,
\be
\label{eq:main_source_accumulation}
\sqrt{\mathbb E\!\left[\left(\sum_{j=1}^{i}(1+r_j)b_{f,j}^{(\#)}\right)^2\right]}
\leq(1+\mathcal S_k)
\begin{cases}
i\lambda_\#,&\#\in\{\overline{\rm BCH},{\rm E}\},\\
\sqrt{i\lambda_\#},&\#\in\{\widetilde{\rm BCH},\Xi\}.
\end{cases}
\ee
Inserting $B$ in each trace gives the same bound multiplied by $\|B\|_{\rm op}$. The drift can therefore accumulate linearly in the number of steps, while the root second moment of each centered sum grows at most as the square root of that number. Sources of different types can be correlated at the same step, so we combine their sums using Minkowski's inequality as follows:
\begin{subequations}
\label{eq:main_combined_source_accumulation}
\begin{align}
\label{eq:bound_combine}
&\sqrt{\mathbb E\!\left[\left(\sum_\#\sum_{j=1}^{i}(1+r_j)b_{f,j}^{(\#)}\right)^2\right]}
\leq(1+\mathcal S_k)\Lambda_i; 
\\
\label{eq:def_cap_lambda}
&\qquad \Lambda_i\equiv \left[i(\lambda_{\overline{\rm BCH}}+\lambda_{\rm E})+\sqrt{i\lambda_{\widetilde{\rm BCH}}}+\sqrt{i\lambda_\Xi}\right].
\end{align}
\end{subequations}
By the constituent relation, Eq.~\eqref{eq:relate_rsb_ideal}, the left-hand side of Eq.~\eqref{eq:main_combined_source_accumulation} equals $\sqrt{\mathbb E[((1+r_i)s_{f,i}-1)^2]}$. Since $\Lambda_i \leq \Lambda_k$ for $i\leq k$, maximizing over $0\leq i\leq f\leq k$ and referring to the definition of $\mathcal{S}_{k}$ in Eq.~\eqref{eq:main_normalization_moment} therefore gives the closed inequality
\be
\label{eq:main_normalization_closure}
\mathcal S_k\leq\Lambda_k(1+\mathcal S_k)
\ee
Assuming $\Lambda_k<1$, this gives the bound
\be
\mathcal S_k\leq\frac{\Lambda_k}{1-\Lambda_k},\qquad 1+\mathcal S_k\leq\frac1{1-\Lambda_k}.
\ee
In particular, when $\Lambda_k<1/2$, we have $\mathcal S_k<1$, which allows us to convert weighted moments into observable fluctuations below.

\section{Bounds on observable bias}\label{sec:supp_gate_bias}
For a Hermitian observable, we define the observable deviation with the corresponding bias and variance as
\be
\label{eq:def_deltao_supp}
\delta O_k=\tn{Tr}[O(\rho_{\psi_k}-\rho_{\phi_k})],\qquad b_O(k)=\mathbb E\delta O_k,\qquad y_O(k)=\tn{Var}(\delta O_k).
\ee
It is useful to define the centered observable $\widetilde O_k=O-\tn{Tr}(O\rho_{\phi_k})\mathbb1$, which satisfies $\|\widetilde O_k\|_{\rm op}\leq2\|O\|_{\rm op}$.
The initial bias vanishes. For $k\geq1$, the exact solution in Eq.~\eqref{eq:sol_z_recursive} separates the deviation into the propagated history and the final source as
\be
\label{eq:main_observable_response}
\delta O_k=\widehat{\delta O}_k+\tn{Tr}(\widetilde O_kJ_k);\qquad \widehat{\delta O}_k=\frac{a_k}{1+r_k}=a_k-r_k\widehat{\delta O}_k, \qquad a_k=\tn{Tr}(\widetilde O_kX_k).
\ee
We decompose the history according to the four sources $\#\in\{\overline{\rm BCH},{\rm E},\widetilde{\rm BCH},\Xi\}$ as
\begin{subequations}
\label{eq:main_bias_source_histories}
\begin{align}
\label{eq:main_bias_observable_history}
a_k^{(\#)}&=\sum_{1\leq j<k}(1+r_j)\tn{Tr}(\widetilde O_kG_{k,j}^{\phi}J_j^{(\#)}G_{k,j}^{\phi}),\\
\label{eq:main_bias_normalization_history}
r_k^{(\#)}&=\sum_{1\leq j<k}(1+r_j)b_{k,j}^{(\#)},\qquad a_k=\sum_\#a_k^{(\#)},\qquad r_k=\sum_\#r_k^{(\#)}.
\end{align}
\end{subequations}
These sums contain $k-1$ time steps, while the terminal source $J_k$ remains separate. Combining the source estimates in Eq.~\eqref{eq:main_source_accumulation} with the bound in Eq.~\eqref{eq:main_normalization_closure} gives
\begin{subequations}
\label{eq:main_bias_history_moments}
\begin{align}
\label{eq:main_bias_drift_history}
\sqrt{\mathbb E[(a_k^{(\nu)})^2]}&\leq\frac{(k-1)\lambda_\nu}{1-\Lambda_k}\|\widetilde O_k\|_{\rm op},\qquad \sqrt{\mathbb E[(r_k^{(\nu)})^2]}\leq\frac{(k-1)\lambda_\nu}{1-\Lambda_k},\qquad \nu\in\{\overline{\rm BCH},{\rm E}\},\\
\label{eq:main_bias_centered_history}
\sqrt{\mathbb E[(a_k^{(\mu)})^2]}&\leq\frac{\sqrt{(k-1)\lambda_\mu}}{1-\Lambda_k}\|\widetilde O_k\|_{\rm op},\qquad \sqrt{\mathbb E[(r_k^{(\mu)})^2]}\leq\frac{\sqrt{(k-1)\lambda_\mu}}{1-\Lambda_k},\qquad \mu\in\{\widetilde{\rm BCH},\Xi\}.
\end{align}
\end{subequations}
Equation~\eqref{eq:main_bias_drift_history} follows from Minkowski's inequality. In Eq.~\eqref{eq:main_bias_centered_history}, centered contributions at different injection times are orthogonal due to martingale orthogonality. 

\subsection{General observables}
These established bounds and estimates allow us to prove the following theorem on the bias of general Hermitian observables.

\begin{theorem}[General observable bias]\label{theorem:supp_bch_bias}
For the global Haar gate realization with TFD initialization and $\Lambda_k<1/2$, every Hermitian doubled-register observable satisfies
\be
\label{eq:supp_bch_bias}
|b_O(k)|\lesssim\frac{\Lambda_k}{1-\Lambda_k}\|O\|_{\rm op}.
\ee
\end{theorem}
For $|\theta|<1$, $|\alpha|\|h\|_{\rm op}<1$, $\|h\|_{\rm op}=O(L)$, and $\beta_kN\|h\|_{\rm op}\gg1$, the bound scales as
\be
\label{eq:main_general_bias_scaling}
|b_O(k)|=O\!\left([\beta_k\|h\|_{\rm op}(|\theta|+|\alpha|\|h\|_{\rm op})+\alpha\theta\beta_k\|h\|_{\rm op}^2]\|O\|_{\rm op}\right).
\ee
With $\alpha=\theta/L$, the leading pulse dependence of this general estimate is linear in $|\theta|$.

\begin{proof}
Let $\widehat b_O(k)=\mathbb E\widehat{\delta O}_k$. Taking the expectation of Eq.~\eqref{eq:main_observable_response} gives
\be
\label{eq:main_bias_mean_identity}
\widehat b_O(k)=\sum_{\nu\in\{\overline{\rm BCH},{\rm E}\}}\mathbb E a_k^{(\nu)} -\sum_\#\mathbb E[r_k^{(\#)}\widehat{\delta O}_k].
\ee
The centered numerator contributions have zero mean, but their correlations with the normalization error can contribute to the bias. Using $|\widehat{\delta O}_k|\leq\|\widetilde O_k\|_{\rm op}$ and Cauchy--Schwarz, we obtain
\begin{subequations}
\label{eq:main_general_history_bias}
\begin{align}
\label{eq:main_general_history_cauchy}
|\widehat b_O(k)|&\leq\sum_\nu\sqrt{\mathbb E[(a_k^{(\nu)})^2]}+\|\widetilde O_k\|_{\rm op}\sum_\#\sqrt{\mathbb E[(r_k^{(\#)})^2]}\\
\label{eq:main_general_history_rates}
&\leq\frac{\|\widetilde O_k\|_{\rm op}}{1-\Lambda_k}\left[2(k-1)(\lambda_{\overline{\rm BCH}}+\lambda_{\rm E})+\sqrt{(k-1)\lambda_{\widetilde{\rm BCH}}}+\sqrt{(k-1)\lambda_\Xi}\right].
\end{align}
\end{subequations}
At the final step, $G_{k,k}^{\phi}=\mathbb1$. Both centered sources have zero conditional mean, so the local drift estimates give
\be
\label{eq:main_general_terminal_bias}
|\mathbb E\tn{Tr}(\widetilde O_kJ_k)| \leq(\lambda_{\overline{\rm BCH}}+\lambda_{\rm E}/N)\|\widetilde O_k\|_{\rm op}.
\ee
The factor $1/N$ in the Euler term reflects the absence of propagator amplification at the terminal step. Adding Eqs.~\eqref{eq:main_general_history_bias} and \eqref{eq:main_general_terminal_bias}, and using $\|\widetilde O_k\|_{\rm op}\leq2\|O\|_{\rm op}$, proves Eq.~\eqref{eq:supp_bch_bias}. Substitution of $k=\beta_k(d+1)/(2\alpha\theta)$ gives the stated scaling.
\end{proof}

\begin{corollary}[Bias of the squared Fubini--Study distance]
\label{corollary:gate_fubini_study_bias}
Under the assumptions of Theorem~\ref{theorem:supp_bch_bias}, define the angular Fubini--Study distance and infidelity by
\be
\label{eq:gate_fubini_study_distance}
\begin{aligned}
\theta_k&\equiv\arccos|\langle\phi_k|\psi_k\rangle|,\\
\mathcal{I}_k&\equiv\sin^2\theta_k=1-|\langle\phi_k|\psi_k\rangle|^2.
\end{aligned}
\ee
The bias of the squared distance is its mean, since its reference value is zero, and
\be
\label{eq:gate_fubini_study_bias}
\mathbb E[\theta_k^2]
\leq
\frac{\pi^2}{4}\mathbb E[\mathcal{I}_k]
\lesssim\frac{\Lambda_k}{1-\Lambda_k}.
\ee
\end{corollary}
\begin{proof}
For the deterministic observable $O=\rho_{\phi_k}$, we have $\|O\|_{\rm op}=1$ and $\delta O_k=-\mathcal{I}_k$.  Theorem~\ref{theorem:supp_bch_bias} therefore gives
\be
\label{eq:gate_mean_infidelity_bound}
\mathbb E[\mathcal{I}_k]
=|b_{\rho_{\phi_k}}(k)|
\lesssim\frac{\Lambda_k}{1-\Lambda_k}.
\ee
The claim follows from $\theta_k^2\leq(\pi^2/4)\sin^2\theta_k=(\pi^2/4)\mathcal{I}_k$, valid for $0\leq\theta_k\leq\pi/2$.
\end{proof}

\subsection{Thermal observables}
For thermal state preparation, a stronger bound holds for any Hermitian observable which has support only on the physical system.
\begin{corollary}[Thermal observable bias]\label{corollary:thermal_observable_bias}
For the shadow chain with $O=O_{\rm sys}\otimes\mathbb1_{\rm anc}$, define
\be
\label{eq:thermal_drift_coefficients}
A_\alpha^*=\frac{6|\alpha\theta|\|h\|_{\rm op}}{d+1}(|\theta|+|\alpha|\|h\|_{\rm op})^2, \qquad \lambda_{\overline{\rm BCH}}^*=A_\alpha^*+C_\alpha.
\ee
If the original normalization parameter satisfies $\Lambda_k<1/2$, then
\be
\label{eq:thermal_observable_bias}
|b_O(k)|\leq\frac{4k\|\widetilde O_k\|_{\rm op}}{(1-\Lambda_k)^2} (\lambda_{\overline{\rm BCH}}^*+\lambda_{\rm E}+\lambda_{\widetilde{\rm BCH}}+\lambda_\Xi).
\ee
For $\alpha=\theta/L$, $\|h\|_{\rm op}=O(L)$, $|\theta|<1$, and $|\alpha|\|h\|_{\rm op}<1$, this is $O[\beta_k\|h\|_{\rm op}(|\theta|+|\alpha|\|h\|_{\rm op})^2\|O\|_{\rm op}]$ at a fixed margin from the threshold.
\end{corollary}

\begin{proof}\label{proof:thermal_observable_bias}
We first derive a symmetry of the ensemble of sampled states. The joint unitary $U_t^{\rm sym}=e^{-itH_{\rm sys}}\otimes e^{itH_{\rm sys}^*}$ fixes the maximally entangled initial state and commutes with $h$. Here complex conjugation is taken in the initialization basis. With the four-pulse update defined as $V_P\equiv e^{i\alpha h}e^{i\theta P}e^{-i\alpha h}e^{-i\theta P}$, the gate transforms as $V_{U_t^{\rm sym}P(U_t^{\rm sym})^\dagger}=U_t^{\rm sym}V_P(U_t^{\rm sym})^\dagger$. Conjugating every projector in a history preserves the Haar measure and the Born weights, since each actual input state transforms with its projector. The endpoint distribution is therefore invariant under this conjugation, and tracing out the ancilla gives
\be
\label{eq:main_thermal_reduced_commutation}
[H_{\rm sys},\tn{Tr}_{\rm anc}\mathbb E\rho_{\psi_k}]=0.
\ee
This commutation relation holds for the averaged reduced state and does not require individual trajectories to commute with the Hamiltonian.

Let $\Pi_E=\Pi_E^{\rm sys}\otimes\mathbb1_{\rm anc}$ be the energy projectors and $O_E=\sum_E\Pi_EO\Pi_E$. This redefined observable $O_E$ is block diagonal in the energy eigenbasis and thus commutes with the Green's function. Both the averaged reduced state and the Gibbs reference are block diagonal in energy, so
\be
\label{eq:main_thermal_pinching}
b_O(k)=b_{O_E}(k),\qquad \widetilde O_{E,k}=\sum_E\Pi_E\widetilde O_k\Pi_E,\qquad \|\widetilde O_{E,k}\|_{\rm op}\leq\|\widetilde O_k\|_{\rm op}.
\ee

For $[B,h]=0$, the energy-basis weights in the Born-weighted Haar contractions are real, so the cubic terms in the observable correction cancel, which leads to the improved $A_\alpha^*$ in Eq.~\eqref{eq:thermal_drift_coefficients}. Specifically, for the actual input $\rho_{\psi_i}$, defining $\mathcal E=V_i-\mathbb1-\alpha\theta[P_i,h]$ and $G=G_{f,i+1}^{\phi}$, we have
\be
\label{eq:main_thermal_commuting_contraction}
\left|\mathbb E_i\tn{Tr}\!\left[BG(\mathcal E\rho_{\psi_i}+\rho_{\psi_i}\mathcal E^\dagger)G\right]\right| \leq A_\alpha^*\|B\|_{\rm op}\tn{Tr}(G\rho_{\psi_i}G),\qquad [B,h]=0.
\ee
Applying this bound with $B=\widetilde O_{E,k}$ and $B=\mathbb1$ gives $\lambda_{\overline{\rm BCH}}^*=A_\alpha^*+C_\alpha$, which should replace the original $\lambda_{\overline{\rm BCH}}$ in the drift history bounds of Eq.~\eqref{eq:main_bias_drift_history}. The normalization estimate still uses the original $\Lambda_k$, and the centered-source bounds are unchanged.

The improved bound for thermal observables follows by bounding each centered-source correlation in the normalization correction as a product of two small factors. We now evaluate all quantities in the source decomposition defined in Eq.~\eqref{eq:main_bias_source_histories} with $O_E$. For $\mu\in\{\widetilde{\rm BCH},\Xi\}$, substituting $\widehat{\delta O}_{E,k}=a_k-r_k\widehat{\delta O}_{E,k}$ into the correlation and applying Cauchy--Schwarz gives
\be
\label{eq:main_thermal_correlation}
|\mathbb E[r_k^{(\mu)}\widehat{\delta O}_{E,k}]| \leq\sqrt{\mathbb E[(r_k^{(\mu)})^2]} \left(\sqrt{\mathbb E[a_k^2]}+\|\widetilde O_{E,k}\|_{\rm op}\sqrt{\mathbb E[r_k^2]}\right).
\ee
The source history moments bound both factors. In particular, Minkowski's inequality gives
\beq
\label{eq:main_thermal_total_history_moment}
\sqrt{\mathbb E[a_k^2]}+\|\widetilde O_{E,k}\|_{\rm op}\sqrt{\mathbb E[r_k^2]}&\leq&\frac{2\|\widetilde O_{E,k}\|_{\rm op}}{1-\Lambda_k}\left[(k-1)(\lambda_{\overline{\rm BCH}}^*+\lambda_{\rm E})
+\sqrt{(k-1)\lambda_{\widetilde{\rm BCH}}}+\sqrt{(k-1)\lambda_\Xi}\right].
\eeq
Combining these estimates in Eq.~\eqref{eq:main_thermal_correlation} and Eq.~\eqref{eq:main_thermal_total_history_moment} with the drift terms in Eq.~\eqref{eq:main_bias_mean_identity} yields
\beq
|\widehat b_{O_E}(k)|&\leq&\frac{2\|\widetilde O_{E,k}\|_{\rm op}}{(1-\Lambda_k)^2}\Big\{(k-1)(\lambda_{\overline{\rm BCH}}^*+\lambda_{\rm E})\big[1-\Lambda_k+\sqrt{(k-1)\lambda_{\widetilde{\rm BCH}}}+\sqrt{(k-1)\lambda_\Xi}\big]\nonumber\\
\label{eq:main_thermal_history_unabsorbed}
&&\hspace{30mm}{}+\big(\sqrt{(k-1)\lambda_{\widetilde{\rm BCH}}}+\sqrt{(k-1)\lambda_\Xi}\big)^2\Big\}.
\eeq
The definition of $\Lambda_k$ in Eq.~\eqref{eq:def_cap_lambda} and Cauchy--Schwarz give
\begin{subequations}
\label{eq:main_thermal_absorption_bounds}
\begin{align}
\label{eq:main_thermal_bracket_bound}
1-\Lambda_k+\sqrt{(k-1)\lambda_{\widetilde{\rm BCH}}}+\sqrt{(k-1)\lambda_\Xi}&\leq1-\Lambda_k+\Lambda_{k-1}\leq1,\\
\label{eq:main_thermal_centered_sum_bound}
\left(\sqrt{(k-1)\lambda_{\widetilde{\rm BCH}}}+\sqrt{(k-1)\lambda_\Xi}\right)^2&\leq2(k-1)(\lambda_{\widetilde{\rm BCH}}+\lambda_\Xi).
\end{align}
\end{subequations}
Applying Eq.~\eqref{eq:main_thermal_absorption_bounds} to Eq.~\eqref{eq:main_thermal_history_unabsorbed} gives
\be
\label{eq:main_thermal_history_bound}
|\widehat b_{O_E}(k)|\leq\frac{2(k-1)\|\widetilde O_{E,k}\|_{\rm op}}{(1-\Lambda_k)^2}\left[\lambda_{\overline{\rm BCH}}^*+\lambda_{\rm E}+2(\lambda_{\widetilde{\rm BCH}}+\lambda_\Xi)\right].
\ee

At the terminal step, the centered sources have zero mean and the drift is bounded by $(\lambda_{\overline{\rm BCH}}^*+\lambda_{\rm E}/N)\|\widetilde O_{E,k}\|_{\rm op}$. Adding this contribution to Eq.~\eqref{eq:main_thermal_history_bound} gives
\be
\label{eq:main_thermal_full_bias}
|b_{O_E}(k)|\leq\frac{\|\widetilde O_{E,k}\|_{\rm op}}{(1-\Lambda_k)^2} \left[(2k-1)(\lambda_{\overline{\rm BCH}}^*+\lambda_{\rm E}) +4(k-1)(\lambda_{\widetilde{\rm BCH}}+\lambda_\Xi)\right]
\ee
, which proves Eq.~\eqref{eq:thermal_observable_bias}. 
\end{proof}

\section{Bounds on trajectory variance}\label{sec:supp_gate_variance}
Let us now turn to proving the bound on the trajectory variance of the observable, following a similar strategy to that used in the bias analysis in Sec.~\ref{sec:supp_bch_error} and Sec.~\ref{sec:supp_gate_bias}.
For complex random variables, we use $\tn{Var}(X)=\mathbb E|X-\mathbb EX|^2$. To bound the propagated state fluctuations entering the drift, we define
\be
\label{eq:main_weighted_state_variance}
\mathcal V_f=\max_{0\leq i\leq f'\leq f}\sup_{B\ne0} \frac{\tn{Var}[(1+r_i)\tn{Tr}(BG_{f',i}^{\phi}\rho_{\psi_i}G_{f',i}^{\phi})]}{\|B\|_{\rm op}^2},
\ee
where $B$ is deterministic and may be non-Hermitian. Here, we assume $\Lambda_f<1/2$. The self-consistent inequality for $\mathcal{S}_f$ in Eq.~\eqref{eq:main_normalization_closure} then gives $1+\mathcal S_f\leq(1-\Lambda_f)^{-1}$ and $\mathcal S_f<1$. Since $s_{i,i}=1$, by the definition of $\mathcal{S}_f$ in Eq.~\eqref{eq:main_normalization_moment} we have $\sqrt{\mathbb E[r_i^2]}\leq\mathcal S_f$, and Cauchy--Schwarz yields $\mathbb E(1+r_i)\geq1-\sqrt{\mathbb E[r_i^2]}\geq1-\mathcal S_f>0$.

Taking $f'=i$ and $B=\mathbb1$ in Eq.~\eqref{eq:main_weighted_state_variance} gives $\tn{Var}(r_i)\leq\mathcal V_f$. For any deterministic $B$, choosing $f'=i$ also gives $\tn{Var}[(1+r_i)\tn{Tr}(B\rho_{\psi_i})]\leq\|B\|_{\rm op}^2\mathcal V_f$. Writing $\mathbb E(1+r_i) = (1+r_i) -(r_i-\mathbb Er_i)$, applying the triangle inequality to this splitting and using $|\tn{Tr}(B\rho_{\psi_i})|\leq\|B\|_{\rm op}$ yields
\be
\label{eq:main_unweighted_state_variance}
\sqrt{\tn{Var}[\tn{Tr}(B\rho_{\psi_i})]} \leq\frac{2\|B\|_{\rm op}}{\mathbb E(1+r_i)}\sqrt{\mathcal V_f} \leq\frac{2\|B\|_{\rm op}}{1-\mathcal S_f}\sqrt{\mathcal V_f}.
\ee

\subsection{Controlling parameter for variance}
Let us now derive a controlling parameter for the variance, by analogy with the analysis in Sec.~\ref{sec:supp_gate_bias}. The key is to generalize the constituent relation in Eq.~\eqref{eq:main_weighted_observable_constituent} to an observable-compatible version, defined as
\be
\label{eq:main_weighted_observable_constituent}
(1+r_i)\tn{Tr}(BG_{f',i}^{\phi}\rho_{\psi_i}G_{f',i}^{\phi})-\tn{Tr}(B\rho_{\phi_{f'}})=\sum_{j=1}^{i}(1+r_j)\tn{Tr}(BG_{f',j}^{\phi}J_jG_{f',j}^{\phi}).
\ee
The sum includes the terminal source $J_i$. To bound the variance, we subtract the ensemble mean and decompose $J_j$ into its four sources. As in Sec.~\ref{subsec:local_source_bias}, let us start by deriving an upper bound for the RHS of Eq.~\eqref{eq:main_weighted_observable_constituent} per source per step. For Hermitian $B$ and $0\leq i<f'\leq f$, the centered BCH contribution satisfies
\be
\label{eq:main_centered_drift_variance}
\sqrt{\tn{Var}[(1+r_{i+1})\tn{Tr}(BG_{f',i+1}^{\phi}J_{i+1}^{\overline{\rm BCH}}G_{f',i+1}^{\phi})]}\leq e^{2\eta\Delta_h}\mathcal D_\alpha\frac{5-2\Lambda_f}{1-2\Lambda_f}\|B\|_{\rm op}\sqrt{\mathcal V_f}
\ee
, where the coefficients $\Delta_h$ and $\mathcal D_\alpha$ are defined as
\be
\label{eq:def_gate_variance_feedback}
\Delta_h=\lambda_{\max}(h)-\lambda_{\min}(h);
\qquad
\mathcal D_\alpha=A_\alpha+\frac{2\mathcal B_{\mathcal W}^2}{d+1}
\ee
,\ with $\lambda(h)$ denoting the eigenvalues of $h$. The coefficient $\mathcal D_\alpha$ is obtained by evaluating the Born-weighted Haar contractions, whereas the $e^{2\eta\Delta_h}$ factor comes from a single-step propagation.

Other sources can be bounded using $\tn{Var}[\#] \leq \mathbb{E}[\#^2]$, so that the bounds in Eq.~\eqref{eq:main_source_accumulation} can be used directly.
For a general $B=B_1+iB_2$, the two Hermitian components obey $\|B_a\|_{\rm op}\leq\|B\|_{\rm op}$. Their real variances add, giving at most a factor $\sqrt2$ in the BCH and centered-source bounds. The Euler estimate already holds for arbitrary $B$. Plugging the bounds in Eq.~\eqref{eq:main_source_accumulation} and Eq.~\eqref{eq:main_centered_drift_variance} into the generalized constituent relation in Eq.~\eqref{eq:main_weighted_observable_constituent} and maximizing over the time indices and $B$ therefore gives
\begin{subequations}
\beq
\label{eq:main_variance_closure}
\sqrt{\mathcal V_f}&\leq&\chi_f\sqrt{\mathcal V_f} +\frac{f\lambda_{\rm E}+\sqrt{2f\lambda_{\widetilde{\rm BCH}}}+\sqrt{2f\lambda_\Xi}}{1-\Lambda_f};
\\
\chi_f&\equiv&\sqrt2 f e^{2\eta\Delta_h}\mathcal D_\alpha\frac{5-2\Lambda_f}{1-2\Lambda_f}.
\eeq
\end{subequations}
For $\chi_f<1$, we absorb the first term into the left-hand side. Since $\lambda_\Xi>0$, the definition of $\lambda_\#$ and the fact that $f\lambda_{\rm E}\leq\Lambda_f<1/2$ give
\be
\label{eq:main_euler_variance_absorption}
\frac{\lambda_{\rm E}}{\lambda_\Xi}=\frac{d+2}{2(d+1)}\leq\frac34, \qquad (f\lambda_{\rm E})^2\leq\frac12 f\lambda_{\rm E}\leq\frac38 f\lambda_\Xi.
\ee
Applying Cauchy--Schwarz to the two remaining square-root terms and using $2+(\sqrt2+\sqrt{3/8})^2<25/4$ gives
\be
\label{eq:main_weighted_variance_bound}
\mathcal V_f\leq\frac{25f(\lambda_{\widetilde{\rm BCH}}+\lambda_\Xi)}{4(1-\Lambda_f)^2(1-\chi_f)^2}.
\ee
Finally, $y_O(k)=\tn{Var}[\tn{Tr}(O\rho_{\psi_k})]$ because the reference is deterministic. Equation~\eqref{eq:main_unweighted_state_variance} at $i=f=k$ and $(1-\mathcal S_k)^{-1}\leq(1-\Lambda_k)/(1-2\Lambda_k)$ convert Eq.~\eqref{eq:main_weighted_variance_bound} into the following result.

\begin{theorem}[Trajectory variance]\label{theorem:var_bound_gate}
For the global Haar gate realization with TFD initialization, the assumptions $k\geq1$, $\Lambda_k<1/2$, and $\chi_k<1$ imply, for every Hermitian $O$,
\be
\label{eq:main_gate_variance}
y_O(k)\leq\frac{25k(\lambda_{\widetilde{\rm BCH}}+\lambda_\Xi)}{(1-2\Lambda_k)^2(1-\chi_k)^2}\|O\|_{\rm op}^2.
\ee
For bounded pulses and fixed margins from both thresholds, this has the scaling
\be
\label{eq:main_variance_scaling}
y_O(k)=O\!\left(\frac{\alpha\theta\beta_k\|h\|_{\rm op}^2}{N} [1+(|\theta|+|\alpha|\|h\|_{\rm op})^2] +\frac{\alpha^3\theta^3\beta_k\|h\|_{\rm op}^4}{N^2}\right)\|O\|_{\rm op}^2.
\ee
\end{theorem}
For local Hamiltonians with $\|h\|_{\rm op}=O(L)$ and $\alpha=\theta/L$, the leading term is $O(\theta^2\beta_k L/N)\|O\|_{\rm op}^2$. This exponential suppression with system size gives trajectory typicality.

\subsection{State-distance bounds}
A typicality bound can also be established for the Fubini--Study distance between the actual and reference states as a corollary. 
\begin{corollary}[Variance of the squared Fubini--Study distance]
\label{corollary:gate_fubini_study_variance}
Under the assumptions of Theorem~\ref{theorem:var_bound_gate}, let $\theta_k,\mathcal{I}_k$ be as in Corollary~\ref{corollary:gate_fubini_study_bias}, and set
\be
\label{eq:gate_fubini_study_mean}
v_k\equiv25k\frac{\lambda_{\widetilde{\rm BCH}}+\lambda_\Xi}{(1-2\Lambda_k)^2(1-\chi_k)^2}.
\ee
Given the bound on $\mathbb E\mathcal{I}_k$ in Eq.~\eqref{eq:gate_mean_infidelity_bound}, if $\mathbb E\mathcal{I}_k\leq I_*<1$ for a fixed constant $I_*$, then
\be
\label{eq:gate_fubini_study_variance}
\tn{Var}(\theta_k^2)\leq\frac{\pi^4}{16(1-I_*)^2}\,v_k.
\ee
\end{corollary}
\begin{proof}
Theorem~\ref{theorem:var_bound_gate} with $O=\rho_{\phi_k}$ gives $\tn{Var}(\mathcal{I}_k)\leq v_k$, since $\delta O_k=-\mathcal{I}_k$. Define $g_{\mathrm{FS}}(x)\equiv(\arcsin\sqrt{x})^2$ for $x\in[0,1]$, so that $\theta_k^2=g_{\mathrm{FS}}(\mathcal{I}_k)$. Convexity of $g_{\mathrm{FS}}$ bounds every secant slope through $\mathbb E\mathcal{I}_k$ by $\pi^2/[4(1-\mathbb E\mathcal{I}_k)]$. Hence
\be
\label{eq:main_fubini_study_secant}
\left|g_{\mathrm{FS}}(\mathcal{I}_k)-g_{\mathrm{FS}}(\mathbb E\mathcal{I}_k)\right|\leq\frac{\pi^2}{4(1-I_*)}|\mathcal{I}_k-\mathbb E\mathcal{I}_k|.
\ee
Variance is at most the mean squared deviation from $g_{\mathrm{FS}}(\mathbb E\mathcal{I}_k)$, so
\be
\label{eq:main_fubini_study_variance_proof}
\tn{Var}(\theta_k^2)\leq\frac{\pi^4}{16(1-I_*)^2}\tn{Var}(\mathcal{I}_k)\leq\frac{\pi^4}{16(1-I_*)^2}\,v_k.
\ee
\end{proof}

\subsection{Global Clifford measurements}\label{sec:supp_clifford}
The exact global Haar-random unitary can be replaced by a uniformly sampled global Clifford 3-design~\cite{Zhu2017Clifford3Design}. Here, we briefly sketch the proof \cite{WangChowdhuryShadowGeometry2026}, which keeps the Haar coefficients $A_\alpha,C_\alpha,\widetilde A_\alpha$ and $\lambda_\Xi$ but introduces new coefficients $\widetilde C_\alpha^{\mathrm{Cl}}$ and $\mathcal D_\alpha^{\mathrm{Cl}}$ for the Clifford ensemble.

For uniformly sampled global Clifford unitaries, the snapshot distribution of the pure stabilizer projectors over the set $\mathrm{Stab}_{2L}$ is
\be
\label{eq:main_clifford_born_law}
\Pr(P|\rho_{\psi_{j-1}})=\frac{d}{|\mathrm{Stab}_{2L}|}\tn{Tr}(\rho_{\psi_{j-1}}P),\qquad P\in\mathrm{Stab}_{2L}.
\ee

Following the gate expansion and norm bounds in Sec.~\ref{subsec:local_source_bias}, the contractions for $A_\alpha,C_\alpha,\widetilde A_\alpha$ and $\lambda_\Xi$ require moments only through order three, including the Born weight. The Clifford $3$-design property~\cite{Zhu2017Clifford3Design} therefore gives the same coefficients. The Euler coefficient $\lambda_{\rm E}$ is independent of the measurement ensemble.

For $\widetilde C_\alpha$, we apply the fifth stabilizer moment~\cite{GrossNezamiWalter2021} to the second moment of $\mathcal W\rho_{\psi_{j-1}}\mathcal W^\dagger$. The same coefficient sums and propagation bounds then give
\be
\label{eq:main_clifford_C_tilde}
\widetilde C_\alpha^{\mathrm{Cl}}=\frac{540d^2\mathcal B_{\mathcal W}^4}{(d+1)(d+2)(d+4)(d+8)}
=\frac94\frac{d+3}{d+8}\widetilde C_\alpha
\leq\frac94\widetilde C_\alpha.
\ee
This replaces $\widetilde C_\alpha$ in Eq.~\eqref{eq:main_source_bounds}.

In the conditional BCH drift variance, the linear term gives $A_\alpha$, while the fifth moment gives the rescaled Haar contribution and $150$ additional Pauli contractions, which gives a redefined coefficient
\be
\label{eq:main_clifford_centered_coefficient}
\mathcal D_\alpha^{\mathrm{Cl}}=A_\alpha+\left[\frac{2(d+3)}{(d+1)(d+8)}+\frac{150d^3}{(d+1)(d+2)(d+4)(d+8)}\right]\mathcal B_{\mathcal W}^2
\leq A_\alpha+\frac{152\mathcal B_{\mathcal W}^2}{d+1}.
\ee
This replaces $\mathcal D_\alpha$ in Eq.~\eqref{eq:main_centered_drift_variance} for Clifford trajectories.

Since these coefficients for Clifford 3-designs only differ from those for the exact Haar-random unitary by $O(1)$ factors, the scaling of the bias and variance bounds in Theorem~\ref{theorem:supp_bch_bias} and Theorem~\ref{theorem:var_bound_gate} remains unchanged with an updated certified margin from the thresholds. The same holds for the Fubini--Study distance bounds in Corollary~\ref{corollary:gate_fubini_study_bias} and Corollary~\ref{corollary:gate_fubini_study_variance}. For Corollary~\ref{corollary:thermal_observable_bias}, the Clifford 3-design violates the commutation relation in Eq.~\eqref{eq:main_thermal_reduced_commutation}, so the improved bound in Corollary~\ref{corollary:thermal_observable_bias} does not apply directly. We leave a more detailed analysis of the thermal observable bias for Clifford 3-designs to future work.

\section{Observable estimation from independent histories}\label{sec:supp_mom}
The median-of-means (MOM) estimator applies to independent shadow-chain trajectories using the total snapshot variance. For a traceless Hermitian $O$, a fresh terminal shadow $\widehat\rho_k=(d+1)P_k-\mathbb1$ gives the unbiased estimator $X_O=\tn{Tr}(O\widehat\rho_k)$ of $\mu_O(k)=\tn{Tr}(O \rho_k)$ for the individual state $\rho_{\psi_k}$ at the terminal time step $k$. With the terminal state fixed, the quantum variance of $X_O$ is $v_O^{\rm sh}(\rho_{\psi_k})=\tn{Var}_{P|\rho}
\left[\tn{Tr}\big(O\widehat\rho_k\big)\right]$.

However, since $\rho_{k}$ also depends on the trajectory history, its mean and variance are $\overline\rho_k=\mathbb E\rho_{\psi_k}$ and $y_O(k)$ as introduced in Eq.~\eqref{eq:path_int} in the main text and Eq.~\eqref{eq:def_deltao_supp} in the Supplementary Section~\ref{sec:supp_gate_bias}, respectively.
Thus, the total variance around the mean $\tn{Tr}(O\overline\rho_k)$ {is}
\be
\label{eq:mom_total_variance}
v_O(k)=y_O(k)+\mathbb E[v_O^{\rm sh}(\rho_{\psi_k})]=\frac{d+1}{d+2}[\tn{Tr}(O^2)+2\tn{Tr}(\overline\rho_kO^2)]-\mu_O(k)^2.
\ee
For example, if $O$ is a traceless Pauli operator, then $v_O(k)=d+1-\mu_O(k)^2=O(d)$; the cost of readout with global shadows can therefore remain large even when the trajectories concentrate. However, different trajectories can be combined into the MOM estimator, which leads to the following theorem.

\begin{theorem}[Median-of-means estimation]\label{theorem:mom_shadow_chain}
For $M$ traceless Hermitian observables $O_1,\ldots,O_M$, let $V_*(k)=\max_m v_{O_m}(k)$, where $v_{O_m}(k)$ is the total variance defined in Eq.~\eqref{eq:mom_total_variance}. Divide $N_{\tn{traj}}\equiv nK$ independent shadow-chain trajectories into $K$ groups of $n$ trajectories, with one terminal shadow at step $k$ from each trajectory. For each $O_m$, average the $n$ values $X_{O_m}$ within each group and take the median of the $K$ group means as $\widehat\mu_m^{\rm MOM}$~\cite{HuangKuengPreskill2020}. For $\epsilon>0$ and $0<\delta<1$, these estimators of $\mu_{O_m}(k)=\tn{Tr}(O_m\overline\rho_k)$ satisfy
\be
\label{eq:mom_simultaneous_bound}
\Pr\!\left[\max_m|\widehat\mu_m^{\rm MOM}-\mu_{O_m}(k)|>\epsilon\right]\leq\delta
\ee
provided the positive integers $n,K$ obey $n\geq4V_*(k)/\epsilon^2$ and $K\geq8\log(M/\delta)$, respectively. Relative to the TFD reference state, the error is at most $\epsilon+|b_{O_m}(k)|$ on the same event.
\end{theorem}

The proof follows the standard MOM argument~\cite{HuangKuengPreskill2020}, with the total variance given by Eq.~\eqref{eq:mom_total_variance} as input. The applicability of MOM approach provides a protocol for observable estimation that is complementary to the direct estimation used in the main text. The logarithmic dependence on the number of observables benefits the multi-observable estimation, inherited directly from the original classical shadow tomography of a single unknown state.

\begin{figure*}[htb!]
\centering
\includegraphics[width=16cm]{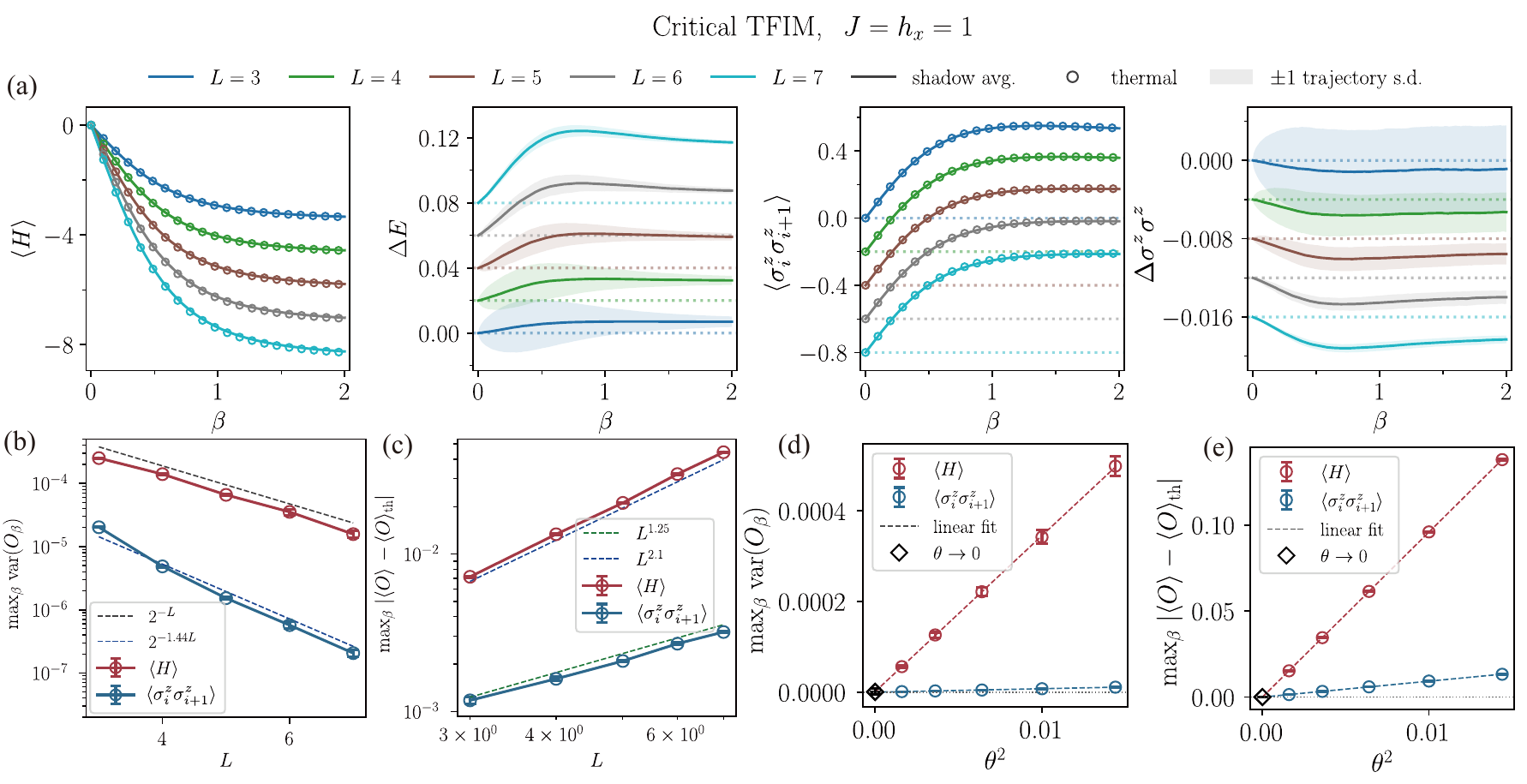}
\caption{\label{Fig:critical_tfim} {\bf Thermal observable estimation using shadow chain for the critical TFIM.} Results are shown for {\bf (a)}-{\bf (c)} system sizes $L=3$--$7$ and $\alpha J=\theta=0.02$, and {\bf (d)}-{\bf (e)} system size $L=5$ and $\alpha = \theta/L$.
{\bf (a)} Energy expectation value and site-averaged nearest-neighbor spin-$Z$ correlation functions with their respective deviations from the exact thermal results. {The color scheme is the same as in Fig.}~\ref{Fig:shadow_result} in the main text. {\bf (b)} Scaling of maximum trajectory variance with the
system size. {\bf (c)} Scaling of maximum cooling bias with the
system size. 
{\bf (d)} {Extrapolation in $\theta$} for the variance.
{\bf (e)} {Extrapolation in $\theta$} for the bias.
}
\end{figure*}

\section{Additional results for spin chains}
\label{sec:supp_additional_benchmarks}

\subsection{Results for other 1D spin-chains}

For completeness, we provide additional benchmarking results on a variety of one-dimensional spin chains for our shadow chain algorithm.

{\it Critical transverse-field Ising model.-} The Hamiltonian of the TFIM tuned to its critical point with open boundary conditions is defined as
\be
H
=
-J\sum_{i=1}^{L-1}\sigma_i^z\sigma_{i+1}^z - J \sum_{i=1}^{L}\sigma_i^x.
\ee
We choose the parameter $J=1$, and the results are shown in Fig.~\ref{Fig:critical_tfim}. They are qualitatively the same as those for the MFIM shown in Fig.~\ref{Fig:shadow_result} of the main text, and are consistent with the proven bounds. This additionally demonstrates that the criticality and integrability tied to this model plays no additional role or constraint on the success of the algorithm.

\begin{figure*}[htb!]
\centering
\includegraphics[width=16cm]{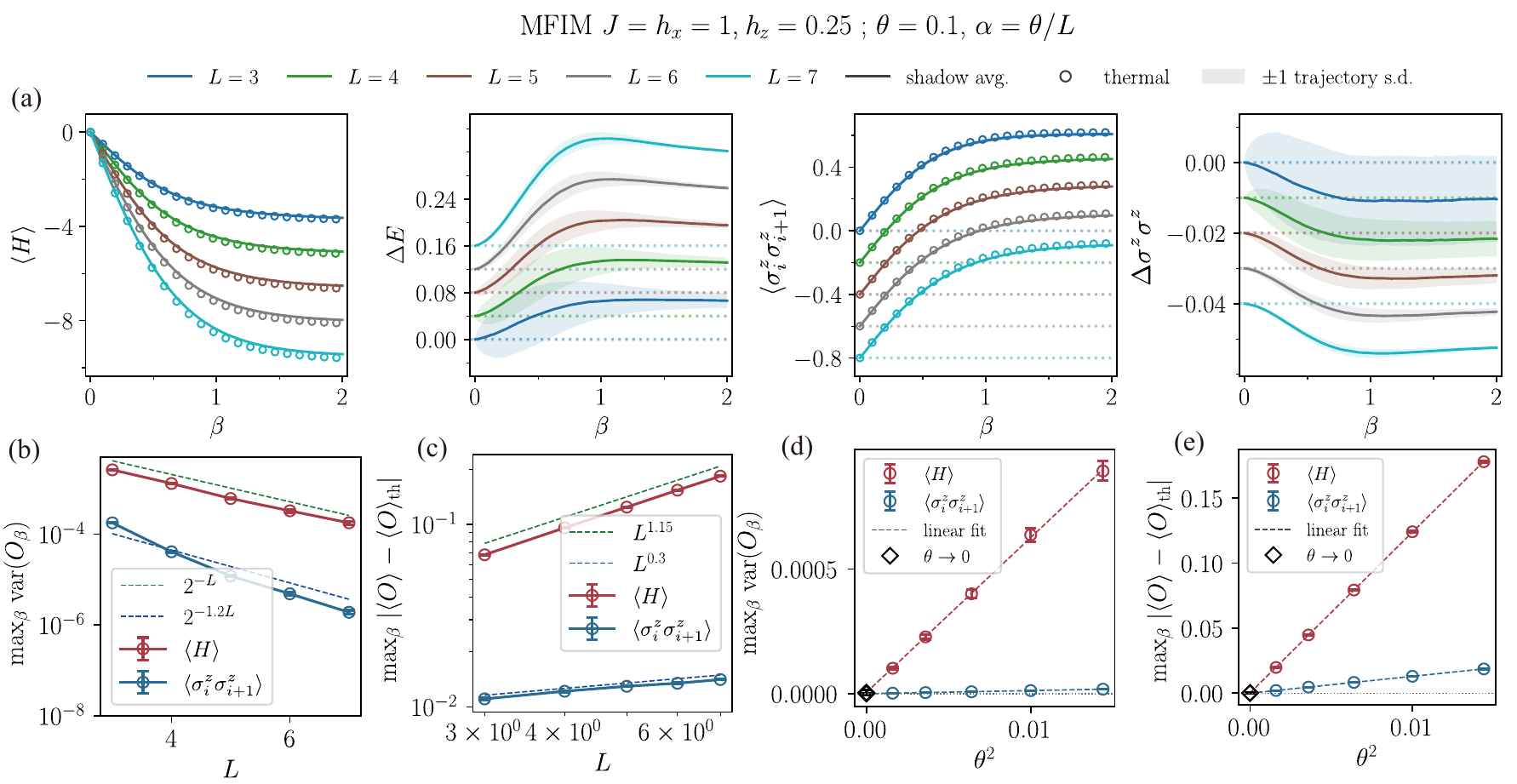}
\caption{\label{Fig:noint_tfim} {\bf Thermal observable estimation using shadow chain for the MFIM.} {Results are shown for system sizes $L=3$--$7$} and $\alpha J L=\theta=0.1$.
{\bf (a)} Energy expectation value and site-averaged nearest-neighbor spin-$Z$ correlation functions {with their respective deviations} from the exact thermal results. {The color scheme is the same as in Fig.}~\ref{Fig:shadow_result} in the main text. {\bf (b)} Scaling of maximum trajectory variance with the
system size. {\bf (c)} Scaling of maximum cooling bias with the
system size. 
{\bf (d)} {Extrapolation in $\theta$} for the variance at $L=5$, with $\alpha = \theta/L$.
{\bf (e)} {Extrapolation in $\theta$} for the bias at $L=5$, with $\alpha = \theta/L$.
}
\end{figure*}

{\it Mixed-field Ising model for different parameter set.-} We revisit the MFIM with open boundary conditions,
\be
H
=
 -J\sum_{i=1}^{L-1}\sigma_i^z\sigma_{i+1}^z - h_x \sum_{i=1}^{L}\sigma_i^x - h_z \sum_{i=1}^{L}\sigma_i^z,
\ee
where we now choose the parameters $J=h_x=1$ and $h_z=0.25$ (i.e. different strength of integrability-breaking term). The results are shown in Fig.~\ref{Fig:noint_tfim}, which are qualitatively similar to Fig.~\ref{Fig:shadow_result} of the main text, and are consistent with the proven bounds.

\begin{figure*}[htb!]
\centering
\includegraphics[width=16cm]{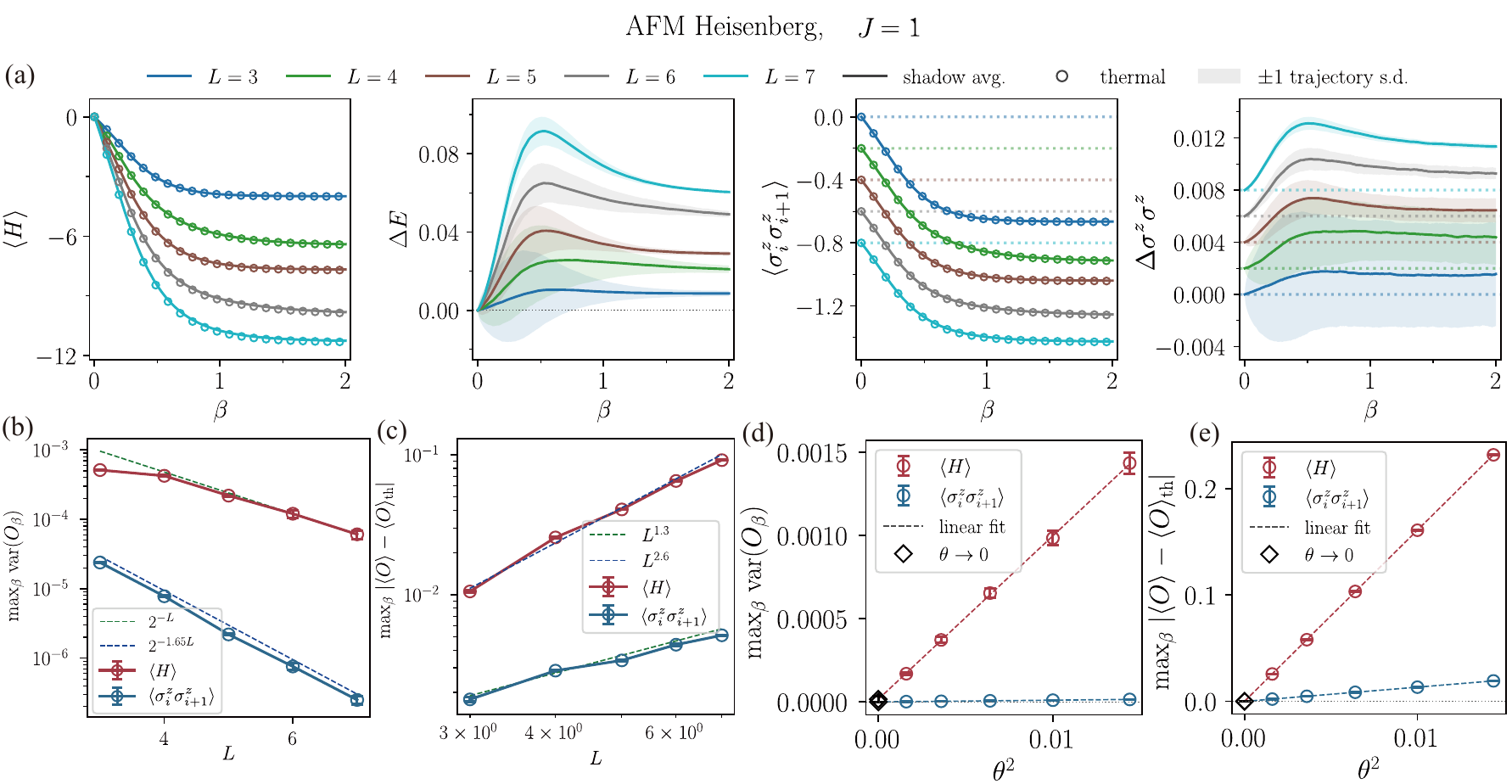}
\caption{\label{Fig:afm_heisenberg} {\bf Thermal observable estimation using shadow chain for the antiferromagnetic Heisenberg chain.} Results are shown for {\bf (a)}-{\bf (c)} system sizes $L=3$--$7$ and $\alpha J=\theta=0.02$, and {\bf (d)}-{\bf (e)} system size $L=5$ and $\alpha = \theta/L$.
{\bf (a)} Energy expectation value and site-averaged nearest-neighbor spin-$Z$ correlation functions {with their respective deviations} from the exact thermal results. {The color scheme is the same as in Fig.}~\ref{Fig:shadow_result} in the main text. {\bf (b)} Scaling of maximum trajectory variance with the
system size. {\bf (c)} Scaling of maximum cooling bias with the
system size. 
{\bf (d)} {Extrapolation in $\theta$} for the variance.
{\bf (e)} {Extrapolation in $\theta$} for the bias.
}
\end{figure*}

{\it Antiferromagnetic Heisenberg chain.-} The Hamiltonian of the antiferromagnetic Heisenberg model (AFMH) with open boundary conditions is defined as
\be
H_{\mathrm{AFMH}}
=
J\sum_{i=1}^{L-1}\left(\sigma_i^x\sigma_{i+1}^x+\sigma_i^y\sigma_{i+1}^y+\sigma_i^z\sigma_{i+1}^z\right).
\ee
These results are qualitatively the same as Fig.~\ref{Fig:shadow_result} of the main text, and are consistent with the proven bounds.

\subsection{Demonstration of the required number of trajectories}

 The results in the main text and the additional results in Fig.~\ref{Fig:critical_tfim}, Fig.~\ref{Fig:noint_tfim} and Fig.~\ref{Fig:afm_heisenberg} focus on extracting the standard deviation among trajectories as a new observable to track its scaling with system size. These extractions require more trajectories than estimating the thermal observables alone. In the main text, we use varying numbers of trajectories $N_{\tn{traj}}$ to obtain satisfactory error bars on the standard deviation itself.

\begin{figure*}[htb!]
\centering
\includegraphics[width=16cm]{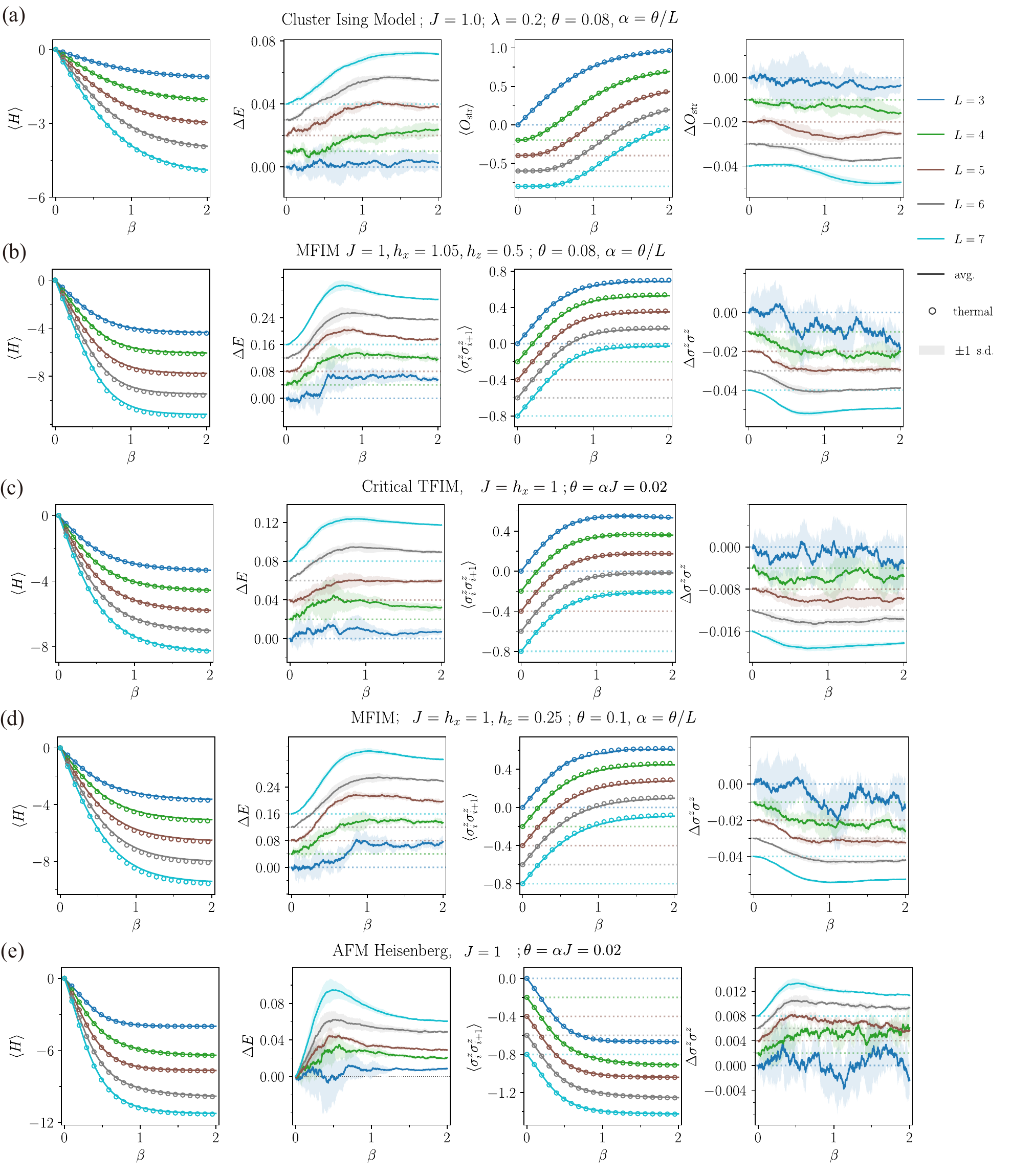}
\caption{\label{Fig:few_traj} {\bf Thermal observable estimation using the shadow chain with a fixed number of trajectories ($N_{\tn{traj}}=4$).} Results are shown for system sizes $L=3$--$7$ for
{\bf (a)} Cluster Ising model, {\bf (b)} Mixed-field Ising model with the same parameters as in the main text, {\bf (c)} Critical transverse-field Ising model, {\bf (d)} Mixed-field Ising model, with the same parameters as in Fig.~\ref{Fig:noint_tfim}, {\bf (e)} Antiferromagnetic Heisenberg model, with the same parameters as in Fig.~\ref{Fig:afm_heisenberg}.
}
\end{figure*}

Here, we show the results for all the benchmarks using a fixed $N_{\tn{traj}}=4$ in Fig.~\ref{Fig:few_traj} as a separate demonstration. The trajectory-averaged results obtained with $N_{\tn{traj}}=4$ using the shadow chain follow the thermal reference results for every benchmark model shown in Fig.~\ref{Fig:few_traj}. This suggests the applicability of the shadow chain protocol with only a few trajectories and gives an estimate of the required number of trajectories for possible experimental realizations.

\end{document}